\documentclass[
 reprint,
 amsmath,amssymb,
 aps,
 pra,
]{revtex4-2}

\usepackage{graphicx}
\usepackage{dcolumn}
\usepackage{bm}
\usepackage{hyperref}
\usepackage{color}

\usepackage{tikz}
\usetikzlibrary{shapes,positioning}

\usetikzlibrary{backgrounds,fit,decorations.pathreplacing,calc}

\makeatletter
\def\grd@save@target#1{
  \def\grd@target{#1}}
\def\grd@save@start#1{
  \def\grd@start{#1}}
\tikzset{
  operator/.style = {draw,fill=white,minimum size=1.5em},
  operator2/.style = {draw,fill=white,minimum height=1.5cm,minimum width=1.75em},
  operator2dash/.style = {draw,dashed,minimum height=2cm,minimum width=3.9em},
  phase/.style = {draw,fill,shape=circle,minimum size=5pt,inner sep=0pt},
  antiphase/.style = {draw,fill=white,shape=circle,minimum size=5pt,inner sep=0pt},
  surround/.style = {fill=blue!10,thick,draw=black,rounded corners=2mm},
  cross/.style={path picture={ 
    \draw[thick,black]
    (path picture bounding box.north) -- (path picture bounding box.south)
    (path picture bounding box.west) -- (path picture bounding box.east);
  }},
  crossx/.style={path picture={ 
    \draw[thick,black,inner sep=0pt]
    (path picture bounding box.south east) -- (path picture bounding box.north west)
    (path picture bounding box.south west) -- (path picture bounding box.north east);
  }},
  circlewc/.style={draw,circle,cross,minimum width=0.5 cm},
  grid with coordinates/.style={
    to path={
      \pgfextra{
        \edef\grd@@target{(\tikztotarget)}
        \tikz@scan@one@point\grd@save@target\grd@@target\relax
        \edef\grd@@start{(\tikztostart)}
        \tikz@scan@one@point\grd@save@start\grd@@start\relax
        \draw[minor help lines] (\tikztostart) grid (\tikztotarget);
        \draw[major help lines] (\tikztostart) grid (\tikztotarget);
        \grd@start
        \pgfmathsetmacro{\grd@xa}{\the\pgf@x/1cm}
        \pgfmathsetmacro{\grd@ya}{\the\pgf@y/1cm}
        \grd@target
        \pgfmathsetmacro{\grd@xb}{\the\pgf@x/1cm}
        \pgfmathsetmacro{\grd@yb}{\the\pgf@y/1cm}
        \pgfmathsetmacro{\grd@xc}{\grd@xa + \pgfkeysvalueof{/tikz/grid with coordinates/major step}}
        \pgfmathsetmacro{\grd@yc}{\grd@ya + \pgfkeysvalueof{/tikz/grid with coordinates/major step}}
        \foreach \x in {\grd@xa,\grd@xc,...,\grd@xb}
        \node[anchor=north] at (\x,\grd@ya) {\pgfmathprintnumber{\x}};
        \foreach \y in {\grd@ya,\grd@yc,...,\grd@yb}
        \node[anchor=east] at (\grd@xa,\y) {\pgfmathprintnumber{\y}};
      }
    }
  },
  minor help lines/.style={
    help lines,
    step=\pgfkeysvalueof{/tikz/grid with coordinates/minor step}
  },
  major help lines/.style={
    help lines,
    line width=\pgfkeysvalueof{/tikz/grid with coordinates/major line width},
    step=\pgfkeysvalueof{/tikz/grid with coordinates/major step}
  },
  grid with coordinates/.cd,
  minor step/.initial=.2,
  major step/.initial=1,
  major line width/.initial=2pt,
}
\makeatother

\usetikzlibrary{fit}
\tikzset{
  highlight/.style={rectangle,fill=lightgray,fill opacity=0.3,inner sep=0pt}
}
\newcommand{\tikzmark}[2]{\tikz[overlay,remember picture,baseline=(#1.base)] \node (#1) {#2};}
\newcommand{\Highlight}[1][submatrix]{
    \tikz[overlay,remember picture]{
    \node[highlight,fit=(left.north west) (right.south east)] (#1) {};}
}

\newcommand{\ket}[1]{\ensuremath{\left\vert#1\right\rangle}}

\begin{document}

\title{Fault-tolerance in qudit circuit design}

\author{Michael~Hanks}
\email{m.hanks@imperial.ac.uk}
\author{M.S.~Kim}
\affiliation{
 QOLS, Blackett Laboratory, Imperial College London, London SW7 2AZ, United Kingdom
}

\date{\today}

\begin{abstract}
    The efficient decomposition of multi-controlled gates is a significant factor in quantum compiling, both in circuit depth and T-gate count.
    Recent work has demonstrated that qudits have the potential to reduce resource requirements from linear to logarithmic depth and to avoid fractional phase rotations.
    Here we argue, based on the scaling of decoherence in high-index states, that circuit depth is not the primary factor, and that both the choice of entangling gate and interaction network topology act together to determine the spread of errors and ultimate failure rate in a circuit.
    We further show that for certain linear-depth circuits, additional error mitigation is possible via selective application of resources.
\end{abstract}

\maketitle

\section{Introduction}
\label{sec:introduction}

Over the past several years, there has been a drive to implement noisy, intermediate-scale quantum (NISQ) algorithms with quantum information processing (QIP) devices of up to 60 qubits
\cite{arute_quantum_2019,wu_strong_2021,zhu_quantum_2022}.
These devices have demonstrated that quantum advantage can be achieved, in principle, with digital devices.
The next major frontier for QIP is operation at scale, extending quantum advantage into the domain of real-world commercial applications.
To achieve this, it is likely that quantum error correction (QEC) will be necessary, but as encoded qubits are still in the early stages of practical development, we may expect the next generation of quantum systems to exist in a hybrid form, with only a subset of critical elements and operations protected via QEC, and supported by a periphery of noisy and perhaps probabilistic devices.
Whatever form the next generation of quantum devices takes, it seems clear that detailed strategies for efficient circuit compilation and low-level error mitigation will retain their significance into the next stage of development.

The decomposition of multi-controlled NOT gates (generalised Toffoli gates) has been a subject of particular interest, as these gates are a frequent source of fractional phase rotations
\cite{barenco_elementary_1995},
and consequently of expensive T-gates in qubit circuits.
Authors have variously pursued strategies involving ancillary qubits
\cite{he_decompositions_2017}
and larger circuit templates
\cite{rahman_templates_2014,rahman_templates_2015,martinez_compiling_2016}.
Notably, linear-depth constructions for multi-controlled gates with constant overhead have been discovered
\cite{maslov_improved_2003,saeedi_linear-depth_2013,gidney_constructing_2015,luo_comment_2016}.
Another approach, seeing recent renewed attention
\cite{gokhale_asymptotic_2019,inada_measurement-free_2021,galda_implementing_2021},
is the use of higher-dimensional spaces --- qudits --- to simplify the decomposition and reduce resources. Here it has been found that a logarithmic depth is possible, and without fractional rotations.

At a glance the qudit approach appears to reduce the problem of multi-controlled Toffoli decomposition almost to one of classical reversible logic; larger dimensions provide space to store results of AND gates that would otherwise require ancillary qubits.
In this paper, we seek to highlight a subtle but important factor in circuit compilation arising from increased sensitivity in higher-index qudit states: state-dependent error rates affect the spread of errors in the circuit in a manner for which gate number and circuit depth cannot account.

In Section~\ref{sec:a_small_example} we demonstrate differential error rates in two na\"{i}vely-equivalent circuit structures. This is followed in Section~\ref{sec:error_scaling_in_large_circuits} by a classification and analysis of several circuit topologies and their relative merits. In Section~\ref{sec:entanglement_induced_error} we discuss the spread of errors due to entangling gates and the key differences between the sensitivity of NISQ quantum devices and QEC-encoded states. Finally, in Section~\ref{sec:discussion} we summarise our observations and discuss implications for near-term QIP applications.

\section{A Small Example}
\label{sec:a_small_example}

Here we describe a small, concrete example to demonstrate that the error in a qudit circuit depends not only on the depth and two-qubit gate count, but also, for multi-qudit gates, on the particular encoding chosen for the parity information (the path through the state space).
We rely on the observation that higher-index states, such as the \ket{2} state, tend to be more sensitive to error than those in the \ket{0/1} subspace.

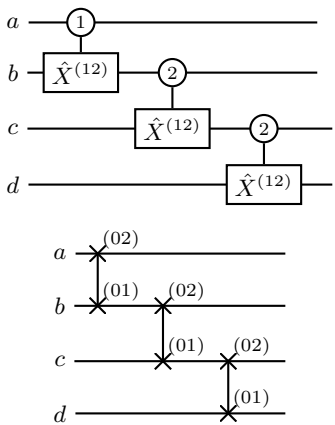
\begin{figure}[ht]
  \begin{center}
  \begin{tabular}{c}
    \begin{tikzpicture}[thick]
      \matrix[row sep=0.2cm, column sep=0.2cm] (circuit) { 
        \node[] (start11) {$a$};
      & \node[circle,fill=white,draw=black,inner sep=0pt,text width=1.0em,align=center] (P11) {\scriptsize$1$};
      &
      &
      & \coordinate (end11);
      \\
        \node[] (start21) {$b$};
      & \node[operator] (X21) {$\hat{X}^{(12)}$};
      & \node[circle,fill=white,draw=black,inner sep=0pt,text width=1.0em,align=center] (P21) {\scriptsize$2$};
      &
      & \coordinate (end21);
      \\
        \node[] (start31) {$c$};
      &
      & \node[operator] (X31) {$\hat{X}^{(12)}$};
      & \node[circle,fill=white,draw=black,inner sep=0pt,text width=1.0em,align=center] (P31) {\scriptsize$2$};
      &
      & \coordinate (end31);
      \\
        \node[] (start41) {$d$};
      &
      &
      & \node[operator] (X41) {$\hat{X}^{(12)}$};
      &
      & \coordinate (end41);
      \\
      };
      \begin{pgfonlayer}{background}
        \draw[thick]
        (start11) -- (end11)
        (start21) -- (end21)
        (start31) -- (end31)
        (start41) -- (end41)
        (P11) -- (X21)
        (P21) -- (X31)
        (P31) -- (X41)
        ;
      \end{pgfonlayer}
    \end{tikzpicture}
    \\
    \begin{tikzpicture}[thick]
      \matrix[row sep=0.2cm, column sep=0.2cm] (circuit) {
        \node[] (start11) {$a$};
      & \node[crossx] (X12) {};
        \node[xshift=0.95em,yshift=0.6em,inner sep=0em] at (X12) {\scriptsize $(02)$};
      &
      &
      & \coordinate (end11);
      \\
        \node[] (start21) {$b$};
      & \node[crossx] (X21) {};
        \node[xshift=0.95em,yshift=0.6em,inner sep=0em] at (X21) {\scriptsize $(01)$};
      & \node[crossx] (X22) {};
        \node[xshift=0.95em,yshift=0.6em,inner sep=0em] at (X22) {\scriptsize $(02)$};
      &
      & \coordinate (end21);
      \\
        \node[] (start31) {$c$};
      &
      & \node[crossx] (X31) {};
        \node[xshift=0.95em,yshift=0.6em,inner sep=0em] at (X31) {\scriptsize $(01)$};
      & \node[crossx] (X32) {};
        \node[xshift=0.95em,yshift=0.6em,inner sep=0em] at (X32) {\scriptsize $(02)$};
      & \coordinate (end31);
      \\
        \node[] (start41) {$d$};
      &
      &
      & \node[crossx] (X41) {};
        \node[xshift=0.95em,yshift=0.6em,inner sep=0em] at (X41) {\scriptsize $(01)$};
      & \coordinate (end41);
      \\
      };
      \begin{pgfonlayer}{background}
        \draw[thick]
        (start11) -- (end11)
        (start21) -- (end21)
        (start31) -- (end31)
        (start41) -- (end41)
        ;
        \draw[thick,shorten >=-4pt,shorten <=-4pt](X12)--(X21);
        \draw[thick,shorten >=-4pt,shorten <=-4pt](X22)--(X31);
        \draw[thick,shorten >=-4pt,shorten <=-4pt](X32)--(X41);
      \end{pgfonlayer}
    \end{tikzpicture}
  \end{tabular}
  \end{center}
  \caption{
    \label{fig:simple_example_extended_linear_circuits}
    The first halves of two linear qutrit circuits for a multi-controlled gate.
    Indices denote control and target sub-spaces, as described in
    Appendix~\ref{sec:qudit_gate_notation}.
    \textbf{(Top)}
    The controlled gate is activated via a rolling succession of controlled excitations.
    \textbf{(Bottom)}
    The controlled gate is deactivated via charge-absorption wherever a qutrit in the $\vert 0\rangle$ state precedes a qutrit in the $\vert 1 \rangle$ state.
  }
\end{figure}

Consider the two sub-circuits shown in
Figure~\ref{fig:simple_example_extended_linear_circuits}.
After the upper (lower) circuit, subsystem $d$ populates the \ket{2} (\ket{1}) state only if all were initially in state \ket{1}.
Both circuits are therefore suitable encoders for a multi-controlled gate, subsequent action depending only on the final state of a single subsystem.
Both circuits involve three, sequential, two-qutrit gates.
That the respective gates can be implemented as elementary operations is seen in \cite{fedorov_implementation_2012} and similar experimental work.

The effect of each circuit in
Figure~\ref{fig:simple_example_extended_linear_circuits} will depend on the initial state.
For the state
\begin{align}
    \left\lvert
        ++++
    \right\rangle
    ,
\end{align}
comprising an even superposition over elements in the computational basis, the result of the upper circuit in
Figure~\ref{fig:simple_example_extended_linear_circuits}
up to normalisation is
\begin{align}
    \vert0000\rangle
    &+
    \vert0001\rangle
    +
    \vert0010\rangle
    +
    \vert0011\rangle
    \nonumber\\
    +
    \vert0100\rangle
    &+
    \vert0101\rangle
    +
    \vert0110\rangle
    +
    \vert0111\rangle
    \nonumber\\
    +
    \vert1000\rangle
    &+
    \vert1001\rangle
    +
    \vert1010\rangle
    +
    \vert1011\rangle
    \nonumber\\
    +
    \vert1\underline{2}00\rangle
    &+
    \vert1\underline{2}01\rangle
    +
    \vert1\underline{22}0\rangle
    +
    \vert1\underline{222}\rangle
    ,
    \label{eq:4_qutrit_excitation_circuit_state}
\end{align}
where we have underlined elements in state \ket{2} for emphasis.
The final state of the lower circuit is
\begin{align}
    \vert0000\rangle
    &+
    \vert00\underline{2}0\rangle
    +
    \vert0\underline{2}00\rangle
    +
    \vert0\underline{22}0\rangle
    \nonumber\\
    +
    \vert\underline{2}000\rangle
    &+
    \vert\underline{2}0\underline{2}0\rangle
    +
    \vert\underline{22}00\rangle
    +
    \vert\underline{222}0\rangle
    \nonumber\\
    +
    \vert1000\rangle
    &+
    \vert10\underline{2}0\rangle
    +
    \vert1\underline{2}00\rangle
    +
    \vert1\underline{22}0\rangle
    \nonumber\\
    +
    \vert1100\rangle
    &+
    \vert11\underline{2}0\rangle
    +
    \vert1110\rangle
    +
    \vert1111\rangle
    .
    \label{eq:4_qutrit_absorption_circuit_state}
\end{align}
It is immediately clear that a larger fraction of the final state for the second circuit populates the \ket{2} component, and this is expected to lead to larger error rates.

\begin{figure}[ht]
  \includegraphics[width=0.49\textwidth]{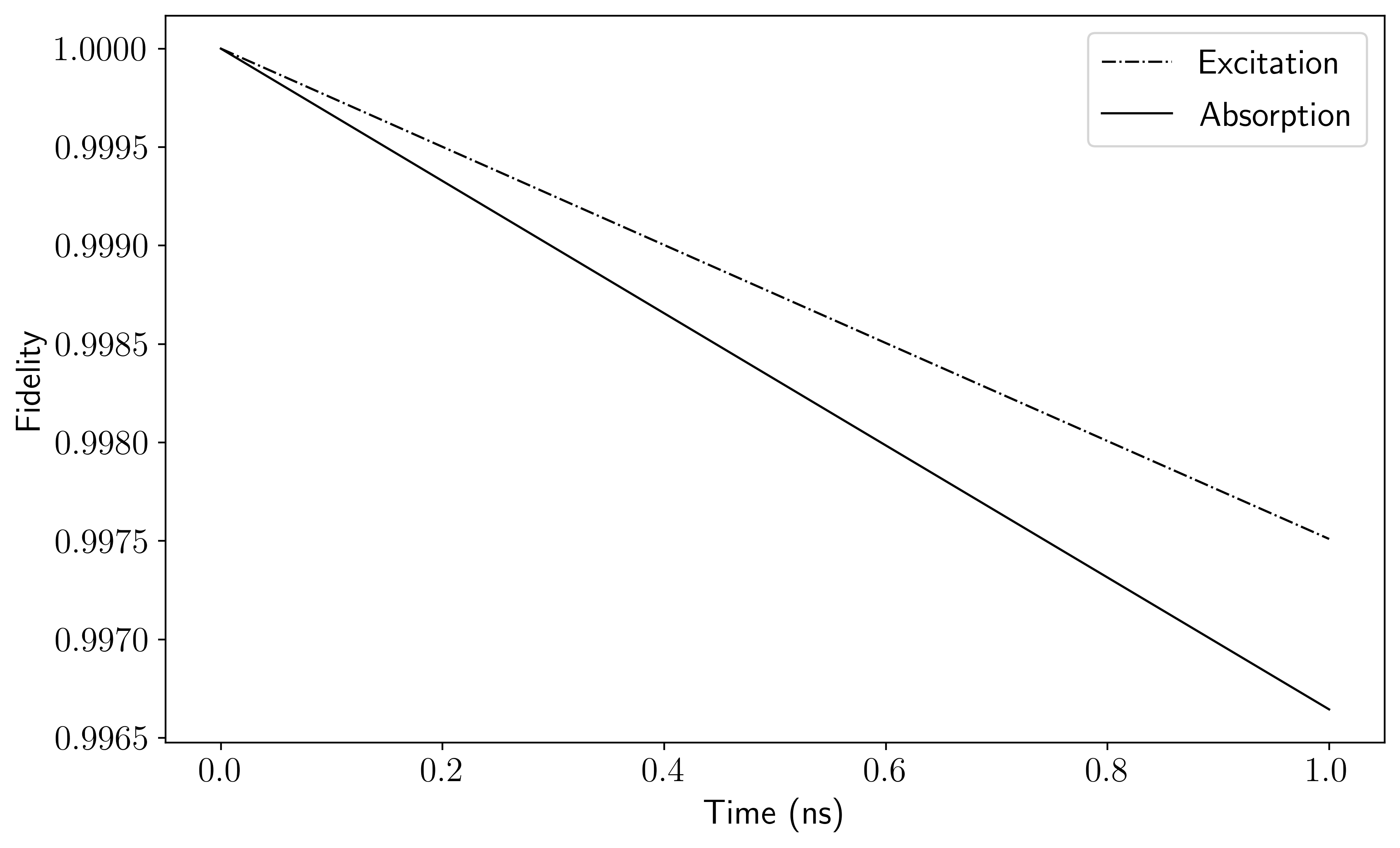}
  \caption{
    \label{fig:spin_chain_linear_decay}
    Reduction in fidelity over time due to decoherence for the final, $4$-qutrit states of the controlled-excitation circuit (Equation~\ref{eq:4_qutrit_excitation_circuit_state}) and the charge-absorption circuit (Equation~\ref{eq:4_qutrit_absorption_circuit_state}).
    Relaxation and dephasing times are drawn from Naik et al. \cite{naik_random_2017} and are given in Equation~\ref{eq:transmon_decay_parameters}.
    Gradients are
    $2.49$~{$\mu$s$^{-1}$}
    (excitation) and
    $3.36$~{$\mu$s$^{-1}$}
    (absorption) respectively.
  }
\end{figure}

As a representative example, take the transmon qutrit of Naik et al. \cite{naik_random_2017}, for which the decoherence times
\begin{align}
    1/\kappa_{01}
    &\sim
    10
    \text{$\mu$s}
    ,\quad
    T_{2,01}
    =
    3.7
    \text{$\mu$s}
    \nonumber\\
    1/\kappa_{12}
    &=
    3.7
    \text{$\mu$s}
    ,\quad
    T^{*}_{2,12}
    =
    1.2
    \text{$\mu$s}
    \label{eq:transmon_decay_parameters}
\end{align}
were reported.
Here the relaxation rate of the \ket{2} state is $2$--$3$~times faster, and the dephasing rate approximately $3$~times faster, than that of the \ket{1} state.
Many qudit systems display similar behaviour.
Using these transmon decay parameters, in
Figure~\ref{fig:spin_chain_linear_decay}
we plot the reduction in fidelity per unit time for each of the final states above,
observing that in this case the charge-absorption circuit is roughly $35$\% more sensitive to decoherence.

The small example above may appear contrived. However, the inferred reduction in fidelity, independent of gate count, is a feature that becomes more significant as the circuit size increases.
In addition to the decoherence rates considered in this example, many qudit interactions themselves involve a state-dependent error distribution, leading to a reduced fidelity for each \textit{active} gate.
These effects depend directly on the topology of the circuit decomposition and on the direction of control (the `flow') of the entangling gates. For the multi-controlled Toffoli gate, in Section~\ref{sec:error_scaling_in_large_circuits} we next explore the available structures and their merits.

\section{Error Scaling in Large Circuits}
\label{sec:error_scaling_in_large_circuits}

In
Section~\ref{sec:a_small_example}
we showed how, in one small instance, the additional error associated with higher-index qudit states can influence the optimal circuit decomposition of multi-controlled gates.
Here we consider larger circuits, with a particular focus on how this additional error scales with the circuit size, relative to the total gate count and circuit depth.

\subsection{Network Taxonomy}
\label{sub:network_taxonomy}

\begin{figure*}[t]
  \begin{center}
  \begin{tabular}{ccc}
    \begin{tikzpicture}[thick]
      \matrix[row sep=0.2cm, column sep=0.2cm] (circuit) {
        \node[] (start11) {$a$};
      & \node[circle,fill=white,draw=black,inner sep=0pt,text width=1.0em,align=center] (P11) {\scriptsize$1$};
      &
      & \node[circle,fill=white,draw=black,inner sep=0pt,text width=1.0em,align=center] (P12) {\scriptsize$1$};
      & \coordinate (end11);
      \\
        \node[] (start21) {$b$};
      & \node[operator] (X21) {$\hat{X}^{(12)}$};
      & \node[circle,fill=white,draw=black,inner sep=0pt,text width=1.0em,align=center] (P21) {\scriptsize$2$};
      & \node[operator] (X22) {$\hat{X}^{(12)}$};
      & \coordinate (end21);
      \\
        \node[] (start31) {$c$};
      &
      & \node[operator] (X31) {$\hat{X}^{(01)}$};
      &
      & \coordinate (end31);
      \\
      };
      \begin{pgfonlayer}{background}
        \draw[thick]
        (start11) -- (end11)
        (start21) -- (end21)
        (start31) -- (end31)
        (P11) -- (X21)
        (P21) -- (X31)
        (P12) -- (X22)
        ;
      \end{pgfonlayer}
    \end{tikzpicture}
    &
    \begin{tikzpicture}[thick]
      \matrix[row sep=0.2cm, column sep=0.2cm] (circuit) {
        \node[] (start11) {$a$};
      & \node[crossx] (X12) {};
        \node[xshift=0.95em,yshift=0.6em,inner sep=0em] at (X12) {\scriptsize $(02)$};
      &
      & \node[crossx] (X13) {};
        \node[xshift=0.95em,yshift=0.6em,inner sep=0em] at (X13) {\scriptsize $(02)$};
      & \coordinate (end11);
      \\
        \node[] (start21) {$b$};
      & \node[crossx] (X21) {};
        \node[xshift=0.95em,yshift=0.6em,inner sep=0em] at (X21) {\scriptsize $(01)$};
      & \node[circle,fill=white,draw=black,inner sep=0pt,text width=1.0em,align=center] (P21) {\scriptsize$1$};
      & \node[crossx] (X22) {};
        \node[xshift=0.95em,yshift=0.6em,inner sep=0em] at (X22) {\scriptsize $(01)$};
      & \coordinate (end21);
      \\
        \node[] (start31) {$c$};
      &
      & \node[operator] (X31) {$\hat{X}^{(01)}$};
      &
      & \coordinate (end31);
      \\
      };
      \begin{pgfonlayer}{background}
        \draw[thick]
        (start11) -- (end11)
        (start21) -- (end21)
        (start31) -- (end31)
        (P21) -- (X31)
        ;
        \draw[thick,shorten >=-4pt,shorten <=-4pt](X12)--(X21);
        \draw[thick,shorten >=-4pt,shorten <=-4pt](X13)--(X22);
      \end{pgfonlayer}
    \end{tikzpicture}
    &
    \begin{tikzpicture}[thick]
      \matrix[row sep=0.2cm, column sep=0.2cm] (circuit) {
        \node[] (start11) {$a$};
      &
      & \node[circle,fill=white,draw=black,inner sep=0pt,text width=1.0em,align=center] (P11) {\scriptsize$0$};
      &
      & \node[circle,fill=white,draw=black,inner sep=0pt,text width=1.0em,align=center] (P12) {\scriptsize$0$};
      &
      & \coordinate (end11);
      \\
        \node[] (start21) {$b$};
      &
      &
      & \node[circle,fill=white,draw=black,inner sep=0pt,text width=1.0em,align=center] (P21) {\scriptsize$1$};
      &
      &
      & \coordinate (end21);
      \\
        \node[] (start31) {$c$};
      & \node[operator] (H31) {$\hat{H}^{(01)}$};
      & \node[operator] (X32) {$\hat{X}^{(12)}$};
      & \node[circle,fill=white,draw=black,inner sep=0pt,text width=1.0em,align=center] (P31) {\scriptsize$1$};
      & \node[operator] (X33) {$\hat{X}^{(12)}$};
      & \node[operator] (H32) {$\hat{H}^{(01)}$};
      & \coordinate (end31);
      \\
      };
      \begin{pgfonlayer}{background}
        \draw[thick]
        (start11) -- (end11)
        (start21) -- (end21)
        (start31) -- (end31)
        (P11) -- (X32)
        (P21) -- (P31)
        (P12) -- (X33)
        ;
      \end{pgfonlayer}
    \end{tikzpicture}
  \end{tabular}
  \end{center}
  \caption{
    \label{fig:network_taxonomy_atoms}
    Three atomic elements in a qutrit decomposition of the multi-controlled Toffoli gate.
    Indices denote control and target sub-spaces, as described in
    Appendix~\ref{sec:qudit_gate_notation}.
    \textbf{(Left)}
    The central gate is activated via controlled excitation of the second qutrit by the first.
    \textbf{(Middle)}
    The central gate is deactivated via charge-absorption from the second qutrit to the first.
    \textbf{(Right)}
    The first qutrit acts directly on the target to de/activate the central gate.
  }
\end{figure*}
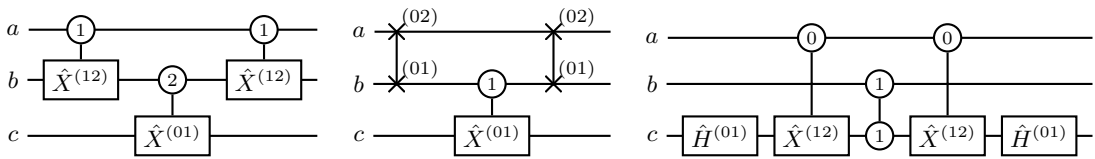

We can divide the atomic elements in a qudit decomposition for the multi-controlled gates into three categories, represented in Figure~\ref{fig:network_taxonomy_atoms}.
In the first two cases,
the first qudit conditionally moves the second \cite{inada_measurement-free_2021} either from an impotent state into a control-subspace \cite{wang_improved_2011,gokhale_asymptotic_2019,galda_implementing_2021} or from the controlling state into an impotent-subspace \cite{fedorov_implementation_2012}.
We will refer to these as \textit{Type-A} and \textit{Type-B} circuits respectively.
As discussed in Section~\ref{sec:a_small_example}, this distinction can have significant consequences.
The third \textit{Type-C} category sees the first qudit act directly on the target \cite{ralph_efficient_2007,lanyon_simplifying_2009}, moving it between affected and invariant sub-spaces.

As the number of control qudits increases, we piece together small sub-circuits of types A through C,
introducing an interaction network topology.
This network topology typically takes the form either of a linear chain \cite{wang_improved_2011}, a log-depth tree \cite{gokhale_asymptotic_2019,inada_measurement-free_2021,galda_implementing_2021}, or a star-like configuration \cite{ralph_efficient_2007,lanyon_simplifying_2009}.
While linear chains and binary trees require only qutrits, the roots of wider branches require larger qudit dimensions.

\subsection{Counting and Errors}
\label{sub:counting_and_errors}

We expect larger error rates to be associated with qudit states of higher index in the computational basis,
and the key problem therefore becomes the counting of these high-index elements.
Consider extending the charge-absorption (Type-B) sub-circuit in
Figure~\ref{fig:simple_example_extended_linear_circuits}
of Section~\ref{sec:a_small_example}
and applying it to the random bit string
\begin{align}
  111\: 011\: 01\: 000\: 011\: 00
  .
\end{align}
In this case the element becomes
\begin{align}
  111\: \textbf{220}\: \textbf{20}\: 000\: \textbf{220}\: 00
  .
\end{align}
Clusters of modified symbols are marked in bold. These clusters, or \textit{runs}, will be our point of interest;
the more widespread and the larger are such clusters in the set of all possible bit-strings, the more sensitive will be intermediate states to decoherence.

The prevalence of modified clusters depends on the initial state as well as the applied circuit.
In
Section~\ref{sec:a_small_example}
we used
$\left\lvert ++++ \right\rangle$
as the initial state.
Since we are interested in the average scaling of the error rate, it is sufficient that initial states be drawn from a distribution such that qudit indices are uncorrelated in the computational basis and evenly distributed between
$\left\lvert 0 \right\rangle$
and
$\left\lvert 1 \right\rangle$.
Given an $n$-qudit circuit and the knowledge that clusters of length $k$ occur with frequency $f\left(k\right)$, we will then infer that the component of the error rate associated with the high-index states scales as
\begin{align}
    S \left( n, f \right)
    &=
    \sum^{n}_{k=1}
    {k} \cdot f\left(k\right)
    ,
\end{align}
and make $S$ our figure of merit.

\subsection{Qutrit Circuits}
\label{sub:qutrit_circuits}

We begin with a restriction to qutrit circuits; the optimal gate count and depth scale linearly and logarithmically in the qudit number regardless of the dimension for $d>2$, and this initial restriction avoids for the moment the question of multiple high-index states.
As mentioned above, qutrits support linear and binary tree topologies.

\subsubsection{Linear Chains}
\label{ssub:linear_chains}

For linear Type-A circuits, high-index states are populated in those bit-strings beginning with a sub-string of $k$ ones. These comprise a fraction $f(k)=2^{-k}$ of the total population of possible computational basis states, and for $n$ control qubits the induced error is therefore expected to scale as
\begin{align}
    S(n, f) &=
  \sum^{n}_{k=1} k 2^{-k}
  \nonumber\\
  &=
  2
  -
  \frac{
    n + 2
  }{
    2^{n}
  }
  .
\end{align}
This approaches the constant $2$ for large $n$, so that the entanglement-induced error can be considered bounded; for this type of circuit, the error associated with higher-index states becomes less important as the circuit size increases and gate-level error models are valid at scale.

For linear Type-B circuits we are no longer tethered to the initial bits; modified clusters may appear in any position, and consequently there is a linear factor increasing the average number of occurrences \emph{per bit}.
The number of $1$-runs of length $k\leq n - 2$ that can be found across all possible binary strings of length $n$ is
$\left(n-k+3\right) 2^{n-k-2}$
\cite{sinha_distribution_2009},
so our frequency distribution in this case should be
\begin{align}
    f\left( k \right)
    &=
    \left(n-k+3\right) 2^{-k-2}
    -
    2^{-k}
    .
\end{align}
The error overhead is therefore expected to scale as
\begin{align}
    S(n, f) &=
    \frac{1}{4}
  \sum^{n}_{k=1}
  k
  \left(n-k\right) 2^{-k}
  \nonumber\\
  &=
  \frac{n}{4}
  \left[
    2
    -
    \frac{
      n + 2
    }{
      2^{n}
    }
  \right]
  -
  \frac{1}{4}
  \left[
    6
    -
    \frac{
      n^{2} + 4n + 6
    }{
      2^{n}
    }
  \right]
  .
\end{align}
This approaches $(n-3)/2$ for large $n$.
Even at scale, the path through state space therefore increases the overall circuit error rate beyond what might be na\"{i}vely assumed with a gate level error model based on isolated one- and two-qubit gates.

The linear chains described here extend the example in
Section~\ref{sec:a_small_example},
confirming that the observed behaviour for the Type-B circuit, rapid decay in the fidelity, is expected to persist at scale.
We next consider binary trees with logarithmic depth.

\subsubsection{Log-Depth Trees}
\label{ssub:log_depth_trees}

Type-A qutrit trees are not possible without either three-qutrit interactions or ancillae:
Information about active gates is stored in the third state of each successive target, so we cannot assign a qutrit as a target in multiple layers without increasing its dimension. We therefore leave a discussion of Type-A circuits to
Section~\ref{sub:higher_dimensions}.

For Type-B binary trees with $n$ qutrits, the number of modified clusters of size $k$ per configuration is expected to be
\begin{align}
    f\left( k \right)
    &=
  n 2^{-2^{k}-k}
  .
\end{align}
A recursive argument for this result is given in the supplementary material.
The error rate is expected to scale as
\begin{align}
    S(n, f)
    &=
    \sum^{n}_{k=1}
    k n 2^{-2^{k}-k}
    \nonumber\\
    &\approx
    \frac{n}{8}
    .
\end{align}
This increases linearly with the number of qudits, just as for the Type-B linear circuit.
However, due to the double-exponential the constant of proportionality is about four times smaller in this case.

So far, for qutrit circuits, we have found that the Type-A linear circuit is the best choice to suppress errors arising from higher-index states. The use of the Type-B binary tree achieves a logarithmic circuit depth, but at the cost of introducing a linear scaling in such error.
We next consider qudits and interactions of higher dimension, to ask whether it is possible to outperform the Type-A linear case, or to strike a compromise.

\subsection{Higher Dimensions}
\label{sub:higher_dimensions}

\subsubsection{Type-A Binary Trees}
\label{ssub:type_a_binary_trees}

\begin{figure*}[t]
  \begin{center}
  \begin{tabular}{cc}
    \begin{tabular}{c}
    \begin{tikzpicture}[thick]
      \matrix[row sep=0.15cm, column sep=0.15cm] (circuit) {
        \node[] (start01) {};
      & \node[circle,fill=white,draw=black,inner sep=0pt,text width=1.0em,align=center] (X01) {\scriptsize$1$};
      &
      &
      & \coordinate (dash0L1);
      &
      & \coordinate (dash0L2);
      &
      & \coordinate (end01);
      \\
        \node[] (start001) {};
      & \node[operator] (X001) {$\hat{X}^{(12)}$};
      & \node[circle,fill=white,draw=black,inner sep=0pt,text width=1.0em,align=center] (X002) {\scriptsize$2$};
      &
      & \coordinate (dash00L1);
      &
      & \coordinate (dash00L2);
      &
      & \coordinate (end001);
      \\
        \node[] (start0001) {};
      & \node[circle,fill=white,draw=black,inner sep=0pt,text width=1.0em,align=center] (X0001) {\scriptsize$1$};
      &
      &
      & \coordinate (dash000L1);
      &
      & \coordinate (dash000L2);
      &
      & \coordinate (end0001);
      \\
        \node[] (start00001) {};
      & \node[operator] (X00001) {$\hat{X}^{(12)}$};
      & \node[operator] (X00002) {$\hat{X}^{(23)}$};
      & \node[circle,fill=white,draw=black,inner sep=0pt,text width=1.0em,align=center] (X00003) {\scriptsize$3$};
      & \coordinate (dash0000L1);
      &
      & \coordinate (dash0000L2);
      &
      & \coordinate (end00001);
      \\
        \node[] (startM1) {};
      &
      &
      & \coordinate (midLeft);
      &
      &
      &
      &
      & \coordinate (endM1);
      \\
        \node[] (startML1) {};
      &
      &
      & \coordinate (midLeftL);
      &
      &
      &
      & \coordinate (midRightL);
      & \coordinate (endML1);
      \\
        \node[] (start11) {};
      & \node[circle,fill=white,draw=black,inner sep=0pt,text width=1.0em,align=center] (X11) {\scriptsize$1$};
      &
      &
      & \coordinate (dash1L1);
      &
      & \coordinate (dash1L2);
      &
      & \coordinate (end11);
      \\
        \node[] (start101) {};
      & \node[operator] (X101) {$\hat{X}^{(12)}$};
      & \node[circle,fill=white,draw=black,inner sep=0pt,text width=1.0em,align=center] (X102) {\scriptsize$2$};
      &
      & \coordinate (dash10L1);
      &
      & \coordinate (dash10L2);
      &
      & \coordinate (end101);
      \\
        \node[] (start1001) {};
      & \node[circle,fill=white,draw=black,inner sep=0pt,text width=1.0em,align=center] (X1001) {\scriptsize$1$};
      &
      &
      & \coordinate (dash100L1);
      &
      & \coordinate (dash100L2);
      &
      & \coordinate (end1001);
      \\
        \node[] (start10001) {};
      & \node[operator] (X10001) {$\hat{X}^{(12)}$};
      & \node[operator] (X10002) {$\hat{X}^{(23)}$};
      & \node[operator] (X10003) {$\hat{X}^{(34)}$};
      & \coordinate (dash1000L1);
      &
      & \coordinate (dash1000L2);
      & \node[operator] (X10004) {$\hat{X}^{(N-1,N)}$};
      & \coordinate (end10001);
      \\
      };
      \begin{pgfonlayer}{background}
        \draw[thick]
        (start01) -- (dash0L1)
        (dash0L2)  -- (end01)
        (start001) -- (dash00L1)
        (dash00L2) -- (end001)
        (start0001) -- (dash000L1)
        (dash000L2) -- (end0001)
        (start00001) -- (dash0000L1)
        (dash0000L2) -- (end00001)
        (start11) -- (dash1L1)
        (dash1L2) -- (end11)
        (start101) -- (dash10L1)
        (dash10L2) -- (end101)
        (start1001) -- (dash100L1)
        (dash100L2) -- (end1001)
        (start10001) -- (dash1000L1)
        (dash1000L2) -- (end10001)
        ;
        \draw[thick,dotted]
        (startM1) -- (endM1)
        (startML1) -- (endML1)
        (dash0L1) -- (dash0L2)
        (dash00L1) -- (dash00L2)
        (dash000L1) -- (dash000L2)
        (dash0000L1) -- (dash0000L2)
        (dash1L1) -- (dash1L2)
        (dash10L1) -- (dash10L2)
        (dash100L1) -- (dash100L2)
        (dash1000L1) -- (dash1000L2)
        ;
        \draw[thick]
        (X01)--(X001)
        (X0001)--(X00001)
        (X002)--(X00002)
        (X11)--(X101)
        (X1001)--(X10001)
        (X102)--(X10002)
        (X00003)--(midLeft)
        (X10003)--(midLeftL)
        (X10004)--(midRightL)
        ;
      \end{pgfonlayer}
    \end{tikzpicture}
    \end{tabular}
    &
    \begin{tabular}{c}
    \begin{tikzpicture}[thick]
      \matrix[row sep=0.15cm, column sep=0.15cm] (circuit) {
        \node[] (start01) {};
      & \node[circle,fill=white,draw=black,inner sep=0pt,text width=1.0em,align=center] (X01) {\scriptsize$1$};
      &
      & \coordinate (dash0L1);
      &
      & \coordinate (dash0L2);
      &
      & \coordinate (end01);
      \\
        \node[] (start001) {};
      & \node[operator] (X001) {$\hat{X}^{(12)}$};
      & \node[circle,fill=white,draw=black,inner sep=0pt,text width=1.0em,align=center] (X002) {\scriptsize$2$};
      & \coordinate (dash00L1);
      &
      & \coordinate (dash00L2);
      &
      & \coordinate (end001);
      \\
        \node[] (start0001) {};
      & \node[circle,fill=white,draw=black,inner sep=0pt,text width=1.0em,align=center] (X0001) {\scriptsize$1$};
      &
      & \coordinate (dash000L1);
      &
      & \coordinate (dash000L2);
      &
      & \coordinate (end0001);
      \\
        \node[] (startM1) {};
      &
      & \coordinate (midLeft);
      &
      &
      &
      &
      & \coordinate (endM1);
      \\
        \node[] (startML1) {};
      &
      & \coordinate (midLeftL);
      &
      &
      &
      & \coordinate (midRightL);
      & \coordinate (endML1);
      \\
        \node[] (start11) {};
      & \node[circle,fill=white,draw=black,inner sep=0pt,text width=1.0em,align=center] (X11) {\scriptsize$1$};
      &
      & \coordinate (dash1L1);
      &
      & \coordinate (dash1L2);
      &
      & \coordinate (end11);
      \\
        \node[] (start101) {};
      & \node[operator] (X101) {$\hat{X}^{(12)}$};
      & \node[circle,fill=white,draw=black,inner sep=0pt,text width=1.0em,align=center] (X102) {\scriptsize$2$};
      & \coordinate (dash10L1);
      &
      & \coordinate (dash10L2);
      &
      & \coordinate (end101);
      \\
        \node[] (start1001) {};
      & \node[circle,fill=white,draw=black,inner sep=0pt,text width=1.0em,align=center] (X1001) {\scriptsize$1$};
      &
      & \coordinate (dash100L1);
      &
      & \coordinate (dash100L2);
      &
      & \coordinate (end1001);
      \\
        \node[] (start100001) {};
      &
      & \node[operator] (X100001) {$\hat{X}^{(12)}$};
      & \coordinate (dash10000L1);
      &
      & \coordinate (dash10000L2);
      &
      & \coordinate (end100001);
      \\
        \node[] (start1000001) {};
      &
      &
      & \coordinate (dash100000L1);
      &
      & \coordinate (dash100000L2);
      & \node[operator] (X1000001) {$\hat{X}^{(12)}$};
      & \coordinate (end1000001);
      \\
      };
      \begin{pgfonlayer}{background}
        \draw[thick]
        (start01) -- (dash0L1)
        (dash0L2)  -- (end01)
        (start001) -- (dash00L1)
        (dash00L2) -- (end001)
        (start0001) -- (dash000L1)
        (dash000L2) -- (end0001)
        (start11) -- (dash1L1)
        (dash1L2) -- (end11)
        (start101) -- (dash10L1)
        (dash10L2) -- (end101)
        (start1001) -- (dash100L1)
        (dash100L2) -- (end1001)
        (start100001) -- (dash10000L1)
        (dash10000L2) -- (end100001)
        (start1000001) -- (dash100000L1)
        (dash100000L2) -- (end1000001)
        ;
        \draw[thick,dotted]
        (startM1) -- (endM1)
        (startML1) -- (endML1)
        (dash0L1) -- (dash0L2)
        (dash00L1) -- (dash00L2)
        (dash000L1) -- (dash000L2)
        (dash1L1) -- (dash1L2)
        (dash10L1) -- (dash10L2)
        (dash100L1) -- (dash100L2)
        (dash10000L1) -- (dash10000L2)
        (dash100000L1) -- (dash100000L2)
        ;
        \draw[thick]
        (X01)--(X001)
        (X0001)--(X001)
        (X11)--(X101)
        (X1001)--(X101)
        (X102)--(midLeftL)
        (X002)--(midLeft)
        (X102)--(X100001)
        (X1000001)--(midRightL)
        ;
      \end{pgfonlayer}
    \end{tikzpicture}
    \end{tabular}
  \end{tabular}
  \end{center}
  \caption{
    \label{fig:type_a_binary_trees}
    Qudit decompositions of the multi-controlled Toffoli gate, using Type-A binary trees.
    Indices denote control and target sub-spaces, as described in
    Appendix~\ref{sec:qudit_gate_notation}.
    \textbf{(Left)}
    Qudits may be `re-used' as targets, but require larger dimensions,
    up to the logarithmic circuit depth.
    \textbf{(Right)}
    Each qudit is a target at most
    once, but `two-controlled' Toffoli gates are required \cite{gokhale_asymptotic_2019}.
  }
\end{figure*}
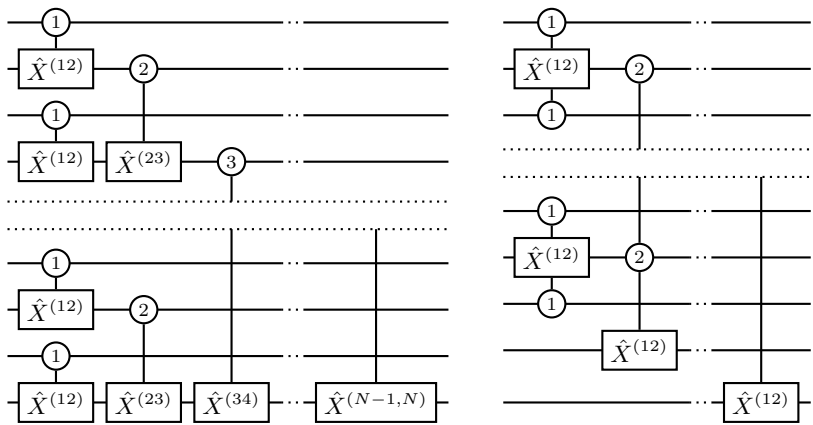

For Type-A trees and without ancillae, there appear to be two methods for implementing binary trees of the first type:
\begin{enumerate}
  \item
  Increase the dimension of each qudit according to the number of times it is assigned as a target, along the lines of the star-like Type-C network topologies
  \cite{ralph_efficient_2007,lanyon_simplifying_2009}.
  \item
  Use two-controlled Toffoli gates as atomic elements \cite{gokhale_asymptotic_2019}, so that outcomes from multiple branches can be combined in a fresh target.
\end{enumerate}
The structures of these circuits are shown in
Figure~\ref{fig:type_a_binary_trees}.

When targets are re-used, modified clusters of size $k$ per configuration are expected with frequency
\begin{align}
    f\left( k \right)
    &=
  \frac{n}{k} 2^{-k}
  .
\end{align}
Recursive arguments for this result and for that of the multi-qudit interaction method below are given in the supplementary material.
Once again, we have exponential decay in $k$, and a linear increase with $n$.
However, in this case larger clusters involve larger state indices, which could exhibit larger error rates.
Specifically, a modified cluster of size $k$ ends with one qudit in state $\left\lvert k/2 \right\rangle$, and $k/2^{r}$ qudits in state $\left\lvert r \right\rangle, 1\leq r<k/2$.
Even in the unlikely case that all high-index states can be treated with the same error rate, this first Type-A tree structure does not outperform the Type-B variant described in
Section~\ref{ssub:log_depth_trees}.

When qudit Toffoli gates are allowed
\cite{gokhale_asymptotic_2019},
modified clusters are expected with frequency
\begin{align}
    f\left( k \right)
    &=
  \frac{n+1}{k+1}
  2^{
    - k
    - \frac{n+1}{k+1}
  }
  .
\end{align}
This is qualitatively different from all other trees considered thus far, as there is also an exponential decay with $\frac{n}{k+1}$.
The network structure in this case is a kind of hybrid between a pure binary tree with independent leaves and the linear chain with all structures tethered to a boundary.
This additional exponential decay bounds the size of the error per configuration as the circuit size increases, so that once again a gate-level error model is sufficient.
Since this tree structure also achieves a logarithmic depth, we might be tempted to conclude that it outperforms its linear counterpart. However, we must first take into account the cost of the additional multi-qudit interaction.
The Type-A tree with three-qudit Toffoli gates uses $(n+1)/2$ such gates. Each gate will involve either a further high-index state or some other novel interaction
\cite{pedersen_native_2019,rasmussen_single-step_2020,roy_programmable_2020,kim_high-fidelity_2021}
that is likely to involve a higher error rate than the more standard single- and two-qudit gates.
The question of whether this circuit outperforms its linear counterpart comes down to whether this additional gate error exceeds the idling error associated with decoherence in the linear-time circuit.

\subsubsection{Star-like Topologies}
\label{ssub:star_like_topologies}

For Type-C circuits, the network extends by increasing the degree of the target node.
Restricting our analysis to single- and two-qudit gates (or more generally for any fixed number), we observe that regardless of the degree of the target node, its interactions proceed linearly as the circuit size becomes large.
At scale then, the primary effect of using nodes of greater degree is to transfer information from multi-qudit, low-index states to a high-index state on a single qudit.
Whether this proves advantageous will depend on the scaling of the error with increasing index (is it better to have two qudits in state
$\left\lvert 2 \right\rangle$
or one qudit in state
$\left\lvert 3 \right\rangle$?),
but in general we do not expect improved performance.
For the transmon system considered in
Section~\ref{sec:a_small_example},
moving from state
$\left\lvert 1 \right\rangle$
to
$\left\lvert 2 \right\rangle$
roughly tripled the decoherence rate.

One further distinction worth noting is between circuits implemented with flip operators in a qubit sub-space as in Figure~\ref{fig:network_taxonomy_atoms} and those implemented with shift-operators that act as
\begin{align}
  \hat{X}^{(\pm)}
  \ket{j}
  &=
  \ket{j\pm1 \text{ mod } N}
  \label{eq:qudit_gate_bit_shift_definition}
  .
\end{align}
A shift-circuit will modify the target state for any control qubit in the \ket{1} state, while later operators in a flip-circuit are only active if all their predecessors were active also. The former therefore displays a linear increase in the expected error rate with circuit size, while the latter is bounded as for the Type-A linear circuit.
Systems with periodic symmetry, such as those encoded in rotationally-symmetric Bosonic codes
\cite{grimsmo_quantum_2020},
may require shift-type operators. Others, such as the transmon systems considered in
Section~\ref{sec:a_small_example},
have a variably-spaced energy level structure that allows for selective transitions.

\section{Entanglement-Induced Error}
\label{sec:entanglement_induced_error}

We established in
Section~\ref{sec:error_scaling_in_large_circuits}
that the Type-A linear circuit produced the intermediate states most robust against the rapid decoherence at high-index computational basis states.
The Type-A circuit has a second significant property:
The high-index states are concentrated in the initial qudits, this being a precondition of the excitation of the later qudits.
It therefore appears that the overall fidelity can be improved by allocating additional resources to the protection of qudits early in the circuit.
However, the circuit involves entangling gates that can affect circuit-level error in two related ways:
\begin{enumerate}
    \item
    The presence of multi-qudit entangled states can amplify correlated rotations.
    For instance, it is well-known that for the GHZ-states, of which the Bell state
    \begin{align}
        \frac{
            \left\lvert 00 \right\rangle
            +
            \left\lvert 11 \right\rangle
        }{\sqrt{2}}
    \end{align}
    represents the smallest example, the sensitivity of the state increases linearly with the qubit number.
    \item
    Discrete errors occurring at intermediate points in the circuit may propagate to also affect other qudits.
    Examples of error propagation for the Type-A linear circuit with four qudits are shown in
    Figure~\ref{fig:linear_type_a_error_propagation_example}.
\end{enumerate}
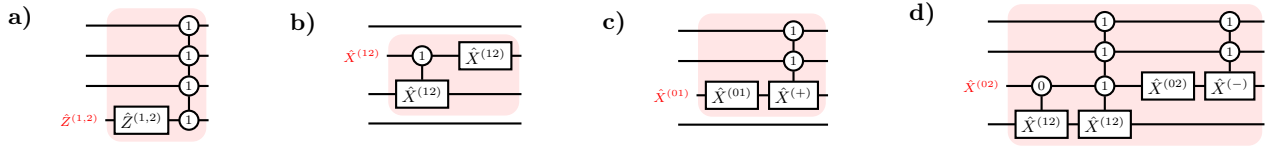
\begin{figure*}[ht!]
  \begin{center}
      \begingroup
      \setlength{\tabcolsep}{1pt}
      \begin{tabular}{cccc}
        \begin{tabular}{rlc}
            \vtop{\null\hbox{\textbf{a)}}}
        &
            \vtop{\null\hbox{
          \begin{tikzpicture}[thick,every node/.style={scale=0.7}]
            \tikzstyle{surround} = [fill=red!10,thick,draw=white,rounded corners=2mm, scale={1.0/0.7}]
            \matrix[row sep=0.12cm, column sep=0.12cm] (circuit) {
              \node[] (start01) {};
            &
            & \node[circle,fill=white,draw=black,inner sep=0pt,text width=1.0em,align=center] (Z03) {\scriptsize$1$};
            & \coordinate (end01);
            \\
              \node[] (start11) {};
            &
            & \node[circle,fill=white,draw=black,inner sep=0pt,text width=1.0em,align=center] (Z17) {\scriptsize$1$};
            & \coordinate (end11);
            \\
              \node[] (start21) {};
            &
            & \node[circle,fill=white,draw=black,inner sep=0pt,text width=1.0em,align=center] (Z29) {\scriptsize$1$};
            & \coordinate (end21);
            \\
              \node[] (start31) {\color{red}\scriptsize$\hat{Z}^{(1,2)}$};
            & \node[operator] (Z37) {$\hat{Z}^{(1,2)}$};
            & \node[circle,fill=white,draw=black,inner sep=0pt,text width=1.0em,align=center] (Z38) {\scriptsize$1$};
            & \coordinate (end31);
            \\
            };
            \begin{pgfonlayer}{background}
              \node[surround] (bg1) [fit = (Z03) (Z37)] {};
              \draw[thick]
              (start01) -- (end01)
              (start11) -- (end11)
              (start21) -- (end21)
              (start31) -- (end31)
              (Z03) -- (Z17)
              (Z17) -- (Z29)
              (Z38) -- (Z29)
              ;
            \end{pgfonlayer}
          \end{tikzpicture}
          }}
          &
          \hspace{2em}
        \end{tabular}
        &
        \begin{tabular}{rlc}
        \vtop{\null\hbox{\textbf{b)}}}
        &
            \vtop{\null\hbox{
          \begin{tikzpicture}[thick,every node/.style={scale=0.7}]
            \tikzstyle{surround} = [fill=red!10,thick,draw=white,rounded corners=2mm, scale={1.0/0.7}]
            \matrix[row sep=0.12cm, column sep=0.12cm] (circuit) {
              \node[] (start01) {};
            &
            &
            & \coordinate (end01);
            \\
              \node[] (start11) {\color{red}\scriptsize$\hat{X}^{(12)}$};
            & \node[circle,fill=white,draw=black,inner sep=0pt,text width=1.0em,align=center] (P12) {\scriptsize$1$};
            & \node[operator] (X13) {$\hat{X}^{(12)}$};
            & \coordinate (end11);
            \\
              \node[] (start21) {};
            & \node[operator] (X24) {$\hat{X}^{(12)}$};
            &
            & \coordinate (end21);
            \\
              \node[] (start31) {};
            &
            &
            & \coordinate (end31);
            \\
            };
            \begin{pgfonlayer}{background}
              \node[surround] (bg1) [fit = (X13) (P12) (X24)] {};
              \draw[thick]
              (start01) -- (end01)
              (start11) -- (end11)
              (start21) -- (end21)
              (start31) -- (end31)
              (P12) -- (X24)
              ;
            \end{pgfonlayer}
          \end{tikzpicture}
          }}
          &
          \hspace{2em}
        \end{tabular}
        &
        \begin{tabular}{rlc}
        \vtop{\null\hbox{\textbf{c)}}}
        &
            \vtop{\null\hbox{
          \begin{tikzpicture}[thick,every node/.style={scale=0.7}]
            \tikzstyle{surround} = [fill=red!10,thick,draw=white,rounded corners=2mm, scale={1.0/0.7}]
            \matrix[row sep=0.12cm, column sep=0.12cm] (circuit) {
              \node[] (start01) {};
            &
            & \node[circle,fill=white,draw=black,inner sep=0pt,text width=1.0em,align=center] (Z03) {\scriptsize$1$};
            & \coordinate (end01);
            \\
              \node[] (start11) {};
            &
            & \node[circle,fill=white,draw=black,inner sep=0pt,text width=1.0em,align=center] (Z17) {\scriptsize$1$};
            & \coordinate (end11);
            \\
              \node[] (start21) {\color{red}\scriptsize$\hat{X}^{(01)}$};
            & \node[operator] (E20) {$\hat{X}^{(01)}$};
            & \node[operator] (E21) {$\hat{X}^{(+)}$};
            & \coordinate (end21);
            \\
              \node[] (start31) {};
            &
            &
            & \coordinate (end31);
            \\
            };
            \begin{pgfonlayer}{background}
              \node[surround] (bg1) [fit = (Z03) (E20) (E21)] {};
              \draw[thick]
              (start01) -- (end01)
              (start11) -- (end11)
              (start21) -- (end21)
              (start31) -- (end31)
              (Z03) -- (Z17)
              (Z17) -- (E21)
              ;
            \end{pgfonlayer}
          \end{tikzpicture}
          }}
          &
          \hspace{2em}
        \end{tabular}
        &
        \begin{tabular}{rlc}
        \vtop{\null\hbox{\textbf{d)}}}
        &
            \vtop{\null\hbox{
          \begin{tikzpicture}[thick,every node/.style={scale=0.7}]
            \tikzstyle{surround} = [fill=red!10,thick,draw=white,rounded corners=2mm, scale={1.0/0.7}]
            \matrix[row sep=0.12cm, column sep=0.12cm] (circuit) {
              \node[] (start01) {};
            &
            & \node[circle,fill=white,draw=black,inner sep=0pt,text width=1.0em,align=center] (E06) {\scriptsize$1$};
            &
            & \node[circle,fill=white,draw=black,inner sep=0pt,text width=1.0em,align=center] (E00) {\scriptsize$1$};
            & \coordinate (end01);
            \\
              \node[] (start11) {};
            &
            & \node[circle,fill=white,draw=black,inner sep=0pt,text width=1.0em,align=center] (E114) {\scriptsize$1$};
            &
            & \node[circle,fill=white,draw=black,inner sep=0pt,text width=1.0em,align=center] (E15) {\scriptsize$1$};
            & \coordinate (end11);
            \\
              \node[] (start21) {\color{red}\scriptsize$\hat{X}^{(02)}$};
            & \node[circle,fill=white,draw=black,inner sep=0pt,text width=1.0em,align=center] (E218) {\scriptsize$0$};
            & \node[circle,fill=white,draw=black,inner sep=0pt,text width=1.0em,align=center] (E224) {\scriptsize$1$};
            & \node[operator] (E211) {$\hat{X}^{(02)}$};
            & \node[operator] (E213) {$\hat{X}^{(-)}$};
            & \coordinate (end21);
            \\
              \node[] (start31) {};
            & \node[operator] (E36) {$\hat{X}^{(12)}$};
            & \node[operator] (E312) {$\hat{X}^{(12)}$};
            &
            &
            & \coordinate (end31);
            \\
            };
            \begin{pgfonlayer}{background}
              \node[surround] (bg1) [fit = (E00) (E06) (E15) (E114) (E211) (E224) (E36) (E312) (E213)] {};
              \draw[thick]
              (start01) -- (end01)
              (start11) -- (end11)
              (start21) -- (end21)
              (start31) -- (end31)
              (E00) -- (E15)
              (E06) -- (E114)
              (E15) -- (E213)
              (E114) -- (E224)
              (E218) -- (E36)
              (E224) -- (E312)
              ;
            \end{pgfonlayer}
          \end{tikzpicture}
          }}
        \end{tabular}
      \end{tabular}
      \setlength{\tabcolsep}{6pt}
      \endgroup
  \end{center}
  \caption{
    \label{fig:linear_type_a_error_propagation_example}
    Propagation of error in the Type-A linear qutrit decomposition.
    Indices denote control and target sub-spaces, as described in
    Appendix~\ref{sec:qudit_gate_notation}.
    Errors are shown to the left of each circuit in red, and are applied in the central portion of the circuit prior to decoding.
    \textbf{a)}
        $\hat{Z}^{(1)}$ and $\hat{Z}^{(2)}$,
    \textbf{b)}
        $\hat{X}^{(12)}$,
    \textbf{c)}
        $\hat{X}^{(01)}$,
    \textbf{d)}
        $\hat{X}^{(02)}$
    .
  }
\end{figure*}

Before drawing conclusions about the allocation of resources for error mitigation, we first investigate the entanglement generated by the Type-A linear circuit between each qudit and the remainder of the state.
Qudits will be labelled in the order in which they are affected in this circuit,
$\left\lvert \psi \right\rangle_{i}$
being the initial state of the $i$th qudit.

\subsection{Entanglement per Qudit}
\label{sub:entanglement_per_qudit}

We begin by looking at the level of entanglement between each qudit and the rest of the state, at that point in the circuit immediately preceding the central controlled operation. Our measure, for simplicity, will be the purity of the partial trace, defined as
\begin{align}
    \text{Tr}
    \left[
        \text{Tr}^{2}_{i}
        \left(
            \hat{\rho}
        \right)
    \right]
    ,
\end{align}
where $\text{Tr}_{i} \left( \hat{\rho} \right)$ denotes the partial trace of $\hat{\rho}$ over the states of qutrit $i$.

\begin{figure}[ht!]
  \includegraphics[width=0.45\textwidth]{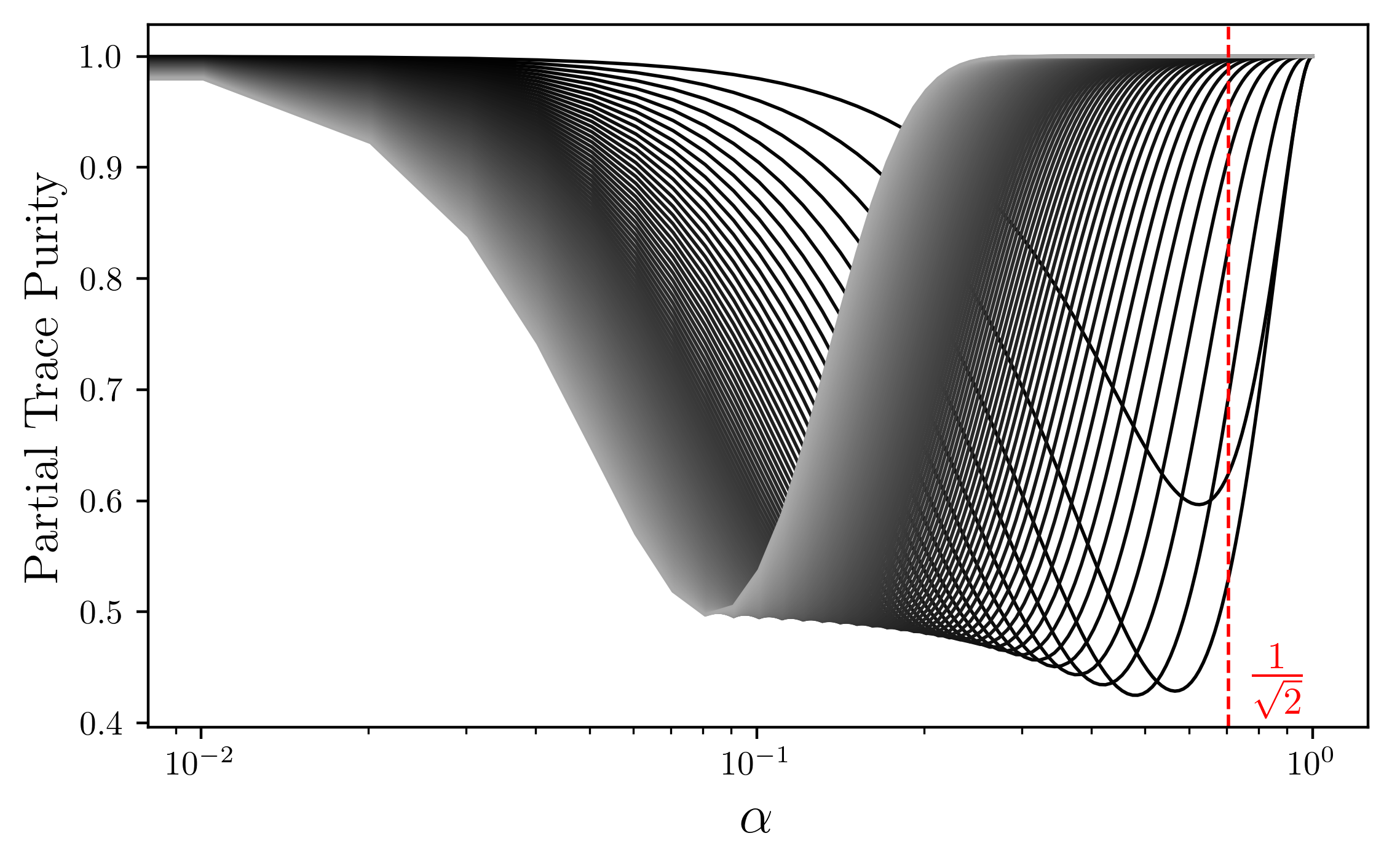}
  \caption{
    \label{fig:type_a_linear_partial_trace_purity}
    The purity of the partial trace for the Type-A linear qutrit circuit. All qutrits are assumed to begin with the same initial state
    $\alpha \vert 0 \rangle + \sqrt{1-\alpha^{2}} \vert 1 \rangle$
    , $\alpha\in\left[0,1\right]$.
    Early qutrits correpond to curves with minima to the right, the qutrit index increasing as the curves move leftward.
  }
\end{figure}

With initial states
\begin{align}
    \vert \psi \rangle_{j}
    &=
    \alpha_{j} \vert 0 \rangle_{j}
    +
    \beta_{j} \vert 1 \rangle_{j}
    ,
\end{align}
the final state for the Type-A linear circuit takes the form
\begin{align}
    \lvert \psi \rangle_{f}
    &=
    \alpha_{0} \vert 0 \rangle_{0}
    \otimes_{j>0} \vert \psi \rangle_{j}
    \nonumber\\
    &\quad
    +
    \beta_{0} \vert 1 \rangle_{0}
    \left(
        \alpha_{1} \vert 0 \rangle_{1}
        \otimes_{j>1} \vert \psi \rangle_{j}
    \right.
    \nonumber\\
    &\qquad
        +
        \beta_{1}  \vert 2 \rangle_{1}
        \left(
            \alpha_{2} \vert 0 \rangle_{2}
            \otimes_{j>2} \vert \psi \rangle_{j}
        \right.
            \nonumber\\
            &\quad\qquad
    \left.
        \left.
            +
            \beta_{2}  \vert 2 \rangle_{2}
            \left(
            \right.
        \right.
    \right.
    \ldots
\end{align}
Noting that
\begin{align}
    \langle 2 \vert \psi \rangle_{j}
    &=
    0
    ,\quad
    \langle 0 \vert \psi \rangle_{j}
    =
    \alpha_{j}
    ,
\end{align}
we find the purity of the partial trace with respect to each qutrit to be
\begin{align}
    \text{Tr}\left(
        \text{Tr}^{2}_{j} \left(
            \hat{\rho}_{f}
        \right)
    \right)
    &=
    \left\lvert \alpha_{j} \right\rvert^{4}
    + \left\lvert \beta_{j} \right\rvert^{4}
    \left[
        P^{2}_{j}
        +
        \left(
            1 - P_{j}
        \right)^{2}
    \right]
    \nonumber\\
    &\quad
    +2 \left\lvert \alpha_{j} \right\rvert^{2} \left\lvert \beta_{j} \right\rvert^{2}
    \left[
        \left\lvert \alpha_{j+1} \right\rvert^{4}
        P^{2}_{j}
        +
        \left(
            1 - P_{j}
        \right)^{2}
    \right]
    \nonumber\\
    P_{j}
    &=
    \prod_{k<j}
    \left\lvert \beta_{k} \right\rvert^{2}
    ,\quad
    P_{0}
    =
    1
    .
    \label{eq:type_a_linear_partial_trace_purity}
\end{align}
For fixed state amplitudes at roughly $2^{-1/2}$ the impurity in the partial trace decays exponentially in $j$, according to $P_{j}$, and we infer that the sensitivity of the state to correlated rotations (increasing at most linearly) is bounded.
Example curves for the partial trace purity are shown in
Figure~\ref{fig:type_a_linear_partial_trace_purity}.

While we have assumed that the initial state is separable, the circuit will not reduce an initial level of entanglement because controlled-gates excite only unoccupied states outside of the initial qubit subspace.

\subsection{Error Dispersion}
\label{sub:error_dispersion}

Our second concern is the propagation of discrete errors occurring mid-way through the circuit.
The Quantum Fisher information might be one valid measure of the sensitivity of the overall state to such errors.
However, as we are interested in the allocation of resources on a qudit-by-qudit basis, we will instead restrict our distance measure to measurements on each individual qudit,
\begin{align}
    D
    \left(
        \hat{\rho}_{f}
        ,
        i
    \right)
    &\equiv
    \left\langle
        \psi
    \right\rvert
    \text{Tr}_{j\neq i}
    \left(
        \hat{\rho}_{f}
    \right)
    \left\lvert
        \psi
    \right\rangle_{i}
    ,
\end{align}
where
$\text{Tr}_{j\neq i} \left( \hat{\rho}_{f} \right)$
indicates the partial trace over all but the $i$th qudit.

Looking first at phase errors,
those of type $\hat{Z}^{(0)}_{k}$ commute with all operators in the circuit, remaining local. Errors of type $\hat{Z}^{(2)}_{k}$ become
\begin{align}
    \hat{I}
    -
    2
    \left(
        \otimes_{j\leq k}
        \vert 1 \rangle
        \langle 1 \vert_{j}
    \right)
    \otimes
    \hat{I}_{>k}
    ,
\end{align}
applying a phase flip only to those states for which all qubits up to and including the target begin in the $\vert 1\rangle$ state. $\hat{Z}^{(1)}_{k}$ errors can be expressed as the product
$-\hat{Z}^{(0)}_{k} \hat{Z}^{(2)}_{k}$.
We therefore need only consider
$\hat{Z}^{(2)}_{k}$, and for this case we find
\begin{align}
    D
    \left(
        \hat{\rho}_{f}
        ,
        i
    \right)
    &=
    1
    -
    4
    P_{i}
    \vert \alpha_{i} \vert^{2}
    \vert \beta_{i} \vert^{2}
    ,
\end{align}
where $P_{i}$ is defined as in
Equation~\ref{eq:type_a_linear_partial_trace_purity}. Once again we have exponential decay in the qutrit index through $P_{i}$, so that the dispersion is bounded at scale.

Turning now to the qutrit generalisation of bit-flip errors, we find that those of type
$\hat{X}^{(12)}_{k}$ affect only the target qutrit and its immediate successor, and are therefore necessarily local. Those of type
$\hat{X}^{(01)}_{k}$ enact this gate on the target, but also apply a cyclic
$\hat{X}^{(+)}_{k}$ shift operation, conditional on all preceding qutrits beginning in state $\vert 1 \rangle$:
\begin{align}
    \hat{I}
    -
    \left(
        \otimes_{j< k}
        \vert 1 \rangle
        \langle 1 \vert_{j}
    \right)
    \otimes
    \hat{I}_{\geq k}
    +
    \left(
        \otimes_{j< k}
        \vert 1 \rangle
        \langle 1 \vert_{j}
    \right)
    \otimes
    \hat{X}^{(+)}_{k}
    \hat{I}_{\geq k}
    .
\end{align}
Finally, bit-flip errors of type
$\hat{X}^{(02)}_{k}$ can be decomposed as
$\hat{X}^{(01)}_{k}\hat{X}^{(12)}_{k}\hat{X}^{(01)}_{k}$. We therefore need only consider $\hat{X}^{(01)}_{k}$, and for this case we find
\begin{align}
    D
    \left(
        \hat{\rho}_{f}
        ,
        i
    \right)
    &=
    1
    -
    P_{i}
    \left(
        1
        -
        \vert \alpha_{i} \vert^{2}
        \vert \beta_{i} \vert^{2}
    \right)
    .
\end{align}
Once again, exponential decay in
$P_{i}$ ensures that error dispersion is bounded at scale.

\subsection{Encoded States}
\label{sub:encoded_states}

As a final point in this section, we briefly mention the effect of quantum error correction codes on the sensitivity of entangled states.

Take the classical repetition code in the
Hadamard basis as an example.
The states are
\begin{align}
    \ket{\pm}_{L}
    &=
    \ket{\pm\pm\pm}
    .
\end{align}
Though the $\ket{\pm}$ states each comprise an evenly weighted superposition between $\ket{0}$ and $\ket{1}$, there nonetheless exist states in superposition,
\begin{align}
    \ket{+++} \pm \ket{---}
    ,
\end{align}
that retain basis states with only even or odd numbers of excitations,
\begin{align}
    \ket{\text{even}}
    &=
    \frac{
        \ket{+++} + \ket{---}
    }{\sqrt{2}}
    \nonumber\\
    &=
    \frac{
          \ket{000}
        + \ket{011}
        + \ket{101}
        + \ket{110}
    }{2}
    ,
    \nonumber\\
    \ket{\text{odd}}
    &=
    \frac{
        \ket{+++} - \ket{---}
    }{\sqrt{2}}
    \nonumber\\
    &=
    \frac{
          \ket{001}
        + \ket{010}
        + \ket{100}
        + \ket{111}
    }{2}
    .
\end{align}
For entangled states of the form
\begin{align}
    \ket{\text{even}} \ket{\text{even}}
    +
    \ket{\text{odd}} \ket{\text{odd}}
    ,
\end{align}
the entanglement induced sensitivity depends on the component of the correlation in the noise channel that lies explicitly along this even/odd axis.
That is, to the extent that the environment is capable of extracting parity information, the encoded states remain vulnerable to the increased sensitivity of entangled states. This is expected to decrease exponentially in the size of the state due to conflicts between transmission and memory time, and as the number of inaccessible qubits (unknown bits of information) increases.

For large encoded states that are biased toward noise in one basis, the width of the code may also be significant. Consider Shor's code as a small example.
The states in this case are extensions of the repetition-encoded states above:
\begin{align}
    \ket{0}_{L}
    &=
    \ket{\text{even}}^{\otimes 3}
    ,\quad
    \ket{1}_{L}
    =
    \ket{\text{odd}}^{\otimes 3}
    .
\end{align}
Entangling two logical qubits now produces
\begin{align}
    \ket{00}_{L} + \ket{11}_{L}
    &=
    \ket{\text{even}}^{\otimes 6}
    +
    \ket{\text{odd}}^{\otimes 6}
    ,
\end{align}
and we find that the sensitivity of the entangled state to even/odd--correlated noise increases linearly with the number of sub-blocks in the code (three, in this case).
Similar arguments can be made for surface code quantum computing, where the parity information lies along any horizontal (resp., vertical) cut of the lattice.

\section{Discussion}
\label{sec:discussion}

Experiments now routinely involve tens of qubits, and these numbers are even readily available via cloud-computing services.
However, current and near-term devices remain heavily restricted by the underlying interaction topology
\cite{nishio_extracting_2020}
and absolute module size
\cite{ionq_inc_future_2021}.
Ancillary qubits exacerbate both of these constraints, and the use of qudit elements therefore appears to be an interesting and efficient alternative.

In this paper, based on the scaling and dispersion of the error expected for high-index states, we argue three points:
\begin{enumerate}
    \item
    State-averaged error rates,
    an assumption for instance of t-designs in randomised benchmarking,
    are insufficient when decomposing large qudit gates.
    \item
    Linear circuits can outperform log-depth trees, even in systems with efficient long-distance interactions
    \cite{baker_memory-equipped_2020}.
    \item
    Local geometric error bounds can be predicted for the purpose of allocating resources for error mitigation, even when entangling gates are present, all qudits and their gate counts are otherwise identical, and there is no post-selection or other error detection.
\end{enumerate}
In this third point certain of the results here are reminiscent of recent work by Hann et al.
\cite{hann_resilience_2021} on the scaling of error in the bucket-brigade QRAM protocol, which focused on restricted structures of entanglement within a larger circuit.
While the error in high-index states is not itself a product of entanglement, the careful design of such structures remains necessary to retain local geometric bounds.

Not all systems suffer equally from increased degradation at higher states. It is easy to imagine for instance that engineered multi-component systems, such as spin chains, cavity arrays and other spatially distinct optical modes would be less vulnerable in this regard, up to local inhomogeneities.
For many popular systems however, including trapped ions, transmon qutrits, and crystal defects, there are significant non-linear dependencies in the energy level structure that give rise to such variation.
Indeed, choosing the most stable and accessible qubit subspace can be a significant element of design in many of these systems.

One avenue for future investigation is the interplay between the scaling observed here and the protection offered by qudit quantum error correction codes, both in their encoding and correction cycles.
Whether qudit-generalisable codes such as the GKP code
\cite{gottesman_encoding_2001}
are capable of varying the redundant space allocated for each state, for instance, and how this interacts with qudit circuit compilation as a resource, would be interesting questions to pursue.

As we have shown, the larger error rates associated with high-index states can impact circuit-level performance with as few as three or four qutrits. Further, the advantage we observe for Type-A linear circuits requires no complicated interaction structure. Particularly in superconducting systems, the third energy level is already commonly used as an additional mechanism for multi-qubit gates
\cite{fedorov_implementation_2012,naik_random_2017}.
We therefore expect our approach and conclusions to be immediately applicable in near-term qudit experiments.

\begin{acknowledgments}
{
    This work is supported by the Samsung GRC grant and the UK Hub in Quantum Computing and Simulation with funding from UKRI EPSRC grant EP/T001062/1.
}
\end{acknowledgments}

\begin{appendix}
\section{Qudit Gate Notation}
\label{sec:qudit_gate_notation}

We are interested in the utility of qudit quantum logic for implementing what are fundamentally qubit entangling gates, so the operators as described in \cite{di_elementary_2012}, which are expressed in terms of respective qubit sub-spaces and Pauli operators, seem most natural.

\begin{figure*}[ht!]
  \begin{center}
    \begin{tikzpicture}[thick]
      \matrix[row sep=0.2cm, column sep=0.2cm] (circuit) {
        \node[] (start11) {};
      & \node[phase] (P11) {};
      & \coordinate (end11);
      &
      & \node[] (start12) {};
      & \node[circle,fill=white,draw=black,inner sep=0pt,text width=1.0em,align=center] (P12) {\scriptsize$1$};
      & \coordinate (end12);
      \\
        \node[] (start21) {};
      & \node[circlewc] (C21) {};
      & \coordinate (end21);
      &
      & \node[] (start22) {};
      & \node[operator] (X21) {$\hat{X}^{(01)}$};
      & \coordinate (end22);
      \\
      };
      \node at ($(end22)+(10pt,0)$){};
      \begin{pgfonlayer}{background}
        \draw[thick]
        (start11) -- (end11)
        (start12) -- (end12)
        (start21) -- (end21)
        (start22) -- (end22)
        (P11) -- (C21)
        (P12) -- (X21)
        ;
        \path[->,thick] ([yshift=-1.1em,xshift=0.2em] end11) edge ([yshift=-1.1em,xshift=0em] start12);
      \end{pgfonlayer}
    \end{tikzpicture}
    \begin{tikzpicture}[thick]
      \matrix[row sep=0.2cm, column sep=0.2cm] (circuit) {
        \node[] (start11) {};
      & \node[phase] (P11) {};
      & \coordinate (end11);
      &
      & \node[] (start12) {};
      & \node[circle,fill=white,draw=black,inner sep=0pt,text width=1.0em,align=center] (P12) {\scriptsize$1$};
      & \coordinate (end12);
      \\
        \node[] (start21) {};
      & \node[phase] (C21) {};
      & \coordinate (end21);
      &
      & \node[] (start22) {};
      & \node[circle,fill=white,draw=black,inner sep=0pt,text width=1.0em,align=center] (X21) {\scriptsize$1$};
      & \coordinate (end22);
      \\
      };
      \node at ($(end22)+(10pt,0)$){};
      \begin{pgfonlayer}{background}
        \draw[thick]
        (start11) -- (end11)
        (start12) -- (end12)
        (start21) -- (end21)
        (start22) -- (end22)
        (P11) -- (C21)
        (P12) -- (X21)
        ;
        \path[->,thick] ([yshift=-1.1em,xshift=0.2em] end11) edge ([yshift=-1.1em,xshift=0em] start12);
      \end{pgfonlayer}
    \end{tikzpicture}
    \begin{tikzpicture}[thick]
      \matrix[row sep=0.2cm, column sep=0.2cm] (circuit) {
        \node[] (start11) {};
      & \node[crossx] (P11) {};
      & \coordinate (end11);
      &
      & \node[] (start12) {};
      & \node[crossx] (P12) {};
        \node[xshift=0.95em,yshift=0.6em,inner sep=0em] at (P12) {\scriptsize $(01)$};
      & \coordinate (end12);
      \\
        \node[] (start21) {};
      & \node[crossx] (C21) {};
      & \coordinate (end21);
      &
      & \node[] (start22) {};
      & \node[crossx] (X21) {};
        \node[xshift=0.95em,yshift=0.6em,inner sep=0em] at (X21) {\scriptsize $(01)$};
      & \coordinate (end22);
      \\
      };
      \node at ($(end22)+(10pt,0)$){};
      \begin{pgfonlayer}{background}
        \draw[thick]
        (start11) -- (end11)
        (start12) -- (end12)
        (start21) -- (end21)
        (start22) -- (end22)
        ;
        \draw[thick,shorten >=-4pt,shorten <=-4pt](P11)--(C21);
        \draw[thick,shorten >=-4pt,shorten <=-4pt](P12)--(X21);
        \path[->,thick] ([yshift=-1.1em,xshift=0.2em] end11) edge ([yshift=-1.1em,xshift=0em] start12);
      \end{pgfonlayer}
    \end{tikzpicture}
  \end{center}
  \caption{
    \label{fig:ternary_controlled_flip_example_circuit}
    Circuit descriptions of three elementary qutrit operations, the controlled
    bit-flip, controlled phase-flip and swap gates, expressed
    as generalisations of equivalent qubit-subspace operations
    \cite{di_elementary_2012}.
    Indices denote control and target sub-spaces, as described in
    Appendix~\ref{sec:qudit_gate_notation}.
  }
\end{figure*}
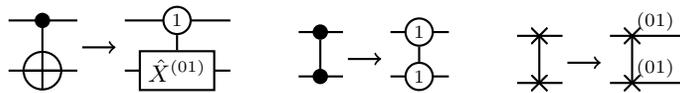

First, the familiar $\hat{X}$, $\hat{Z}$, $\hat{H}$, $\hat{CX}$, $\hat{CZ}$ and similar gates are extended by assigning control-subscript and target-superscript to the operators to denote the computational basis elements involved. For example, we have
\begin{align}
  &
  \hat{X}^{(01)}
  \equiv
  \begin{bmatrix}
    \tikzmark{left}{0} & 1 & 0 \\
    1 & \tikzmark{right}{0} & 0 \\
    0 & 0 & 1 \\
  \end{bmatrix},
  \Highlight[first]
  \quad
  \hat{Z}^{(1)}
  \equiv
  \begin{bmatrix}
    1 & 0 & 0 \\
    0 & -1 & 0 \\
    0 & 0 & 1 \\
  \end{bmatrix},
  \nonumber\\
  &\quad
  \hat{CX}^{(01)}_{(1)}
  \equiv
  \begin{bmatrix}
    1 & 0 & 0 & 0 & 0 & 0 & 0 & 0 & 0 \\
    0 & 1 & 0 & 0 & 0 & 0 & 0 & 0 & 0 \\
    0 & 0 & 1 & 0 & 0 & 0 & 0 & 0 & 0 \\
    0 & 0 & 0 & \tikzmark{left}{0} & 1 & 0 & 0 & 0 & 0 \\
    0 & 0 & 0 & 1 & 0 & 0 & 0 & 0 & 0 \\
    0 & 0 & 0 & 0 & 0 & \tikzmark{right}{1} & 0 & 0 & 0 \\
    0 & 0 & 0 & 0 & 0 & 0 & 1 & 0 & 0 \\
    0 & 0 & 0 & 0 & 0 & 0 & 0 & 1 & 0 \\
    0 & 0 & 0 & 0 & 0 & 0 & 0 & 0 & 1
  \end{bmatrix}
  \Highlight[second]
  \label{eq:qudit_gate_examples}
  .
\end{align}
In circuit diagrams, the filled and empty circles that usually denote control qubits and phase flips are similarly replaced with nodes containing numbered indices, as shown in Figure~\ref{fig:ternary_controlled_flip_example_circuit}.

\end{appendix}


%

\end{document}


\title{Fault-tolerance in qudit circuit design}

\author{Michael~Hanks}
\email{m.hanks@imperial.ac.uk}
\author{M.S.~Kim}
\affiliation{
 QOLS, Blackett Laboratory, Imperial College London, London SW7 2AZ, United Kingdom
}

\date{\today}

\widetext
\begin{center}
\textbf{\large Supplementary Material:
Fault-tolerance in qudit circuit design
}
\end{center}
\setcounter{equation}{0}
\setcounter{figure}{0}
\setcounter{table}{0}
\setcounter{page}{1}
\setcounter{section}{0}
\setcounter{subsection}{0}
\makeatletter
\renewcommand{\theequation}{S\arabic{equation}}
\renewcommand{\thefigure}{S\arabic{figure}}
\renewcommand{\bibnumfmt}[1]{[S#1]}
\renewcommand{\citenumfont}[1]{S#1}

\section{Binary Tree Network Topologies}
\label{sec:binary_tree_network_topologies}

\subsection{Type-B Circuit}
\label{sub:type_1b_circuit}

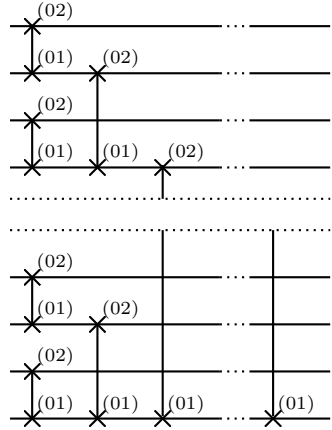
\begin{figure}[ht!]
  \begin{center}
  \begin{tabular}{c}
    \begin{tikzpicture}[thick]
      \matrix[row sep=0.2cm, column sep=0.2cm] (circuit) {
        \node[] (start01) {};
      & \node[crossx] (X01) {};
        \node[xshift=0.95em,yshift=0.6em,inner sep=0em] at (X01) {\scriptsize $(02)$};
      &
      &
      & \coordinate (dash0L1);
      &
      & \coordinate (dash0L2);
      &
      & \coordinate (end01);
      \\
        \node[] (start001) {};
      & \node[crossx] (X001) {};
        \node[xshift=0.95em,yshift=0.6em,inner sep=0em] at (X001) {\scriptsize $(01)$};
      & \node[crossx] (X002) {};
        \node[xshift=0.95em,yshift=0.6em,inner sep=0em] at (X002) {\scriptsize $(02)$};
      &
      & \coordinate (dash00L1);
      &
      & \coordinate (dash00L2);
      &
      & \coordinate (end001);
      \\
        \node[] (start0001) {};
      & \node[crossx] (X0001) {};
        \node[xshift=0.95em,yshift=0.6em,inner sep=0em] at (X0001) {\scriptsize $(02)$};
      &
      &
      & \coordinate (dash000L1);
      &
      & \coordinate (dash000L2);
      &
      & \coordinate (end0001);
      \\
        \node[] (start00001) {};
      & \node[crossx] (X00001) {};
        \node[xshift=0.95em,yshift=0.6em,inner sep=0em] at (X00001) {\scriptsize $(01)$};
      & \node[crossx] (X00002) {};
        \node[xshift=0.95em,yshift=0.6em,inner sep=0em] at (X00002) {\scriptsize $(01)$};
      & \node[crossx] (X00003) {};
        \node[xshift=0.95em,yshift=0.6em,inner sep=0em] at (X00003) {\scriptsize $(02)$};
      & \coordinate (dash0000L1);
      &
      & \coordinate (dash0000L2);
      &
      & \coordinate (end00001);
      \\
        \node[] (startM1) {};
      &
      &
      & \coordinate (midLeft);
      &
      &
      &
      &
      & \coordinate (endM1);
      \\
        \node[] (startML1) {};
      &
      &
      & \coordinate (midLeftL);
      &
      &
      &
      & \coordinate (midRightL);
      & \coordinate (endML1);
      \\
        \node[] (start11) {};
      & \node[crossx] (X11) {};
        \node[xshift=0.95em,yshift=0.6em,inner sep=0em] at (X11) {\scriptsize $(02)$};
      &
      &
      & \coordinate (dash1L1);
      &
      & \coordinate (dash1L2);
      &
      & \coordinate (end11);
      \\
        \node[] (start101) {};
      & \node[crossx] (X101) {};
        \node[xshift=0.95em,yshift=0.6em,inner sep=0em] at (X101) {\scriptsize $(01)$};
      & \node[crossx] (X102) {};
        \node[xshift=0.95em,yshift=0.6em,inner sep=0em] at (X102) {\scriptsize $(02)$};
      &
      & \coordinate (dash10L1);
      &
      & \coordinate (dash10L2);
      &
      & \coordinate (end101);
      \\
        \node[] (start1001) {};
      & \node[crossx] (X1001) {};
        \node[xshift=0.95em,yshift=0.6em,inner sep=0em] at (X1001) {\scriptsize $(02)$};
      &
      &
      & \coordinate (dash100L1);
      &
      & \coordinate (dash100L2);
      &
      & \coordinate (end1001);
      \\
        \node[] (start10001) {};
      & \node[crossx] (X10001) {};
        \node[xshift=0.95em,yshift=0.6em,inner sep=0em] at (X10001) {\scriptsize $(01)$};
      & \node[crossx] (X10002) {};
        \node[xshift=0.95em,yshift=0.6em,inner sep=0em] at (X10002) {\scriptsize $(01)$};
      & \node[crossx] (X10003) {};
        \node[xshift=0.95em,yshift=0.6em,inner sep=0em] at (X10003) {\scriptsize $(01)$};
      & \coordinate (dash1000L1);
      &
      & \coordinate (dash1000L2);
      & \node[crossx] (X10004) {};
        \node[xshift=0.95em,yshift=0.6em,inner sep=0em] at (X10004) {\scriptsize $(01)$};
      & \coordinate (end10001);
      \\
      };
      \begin{pgfonlayer}{background}
        \draw[thick]
        (start01) -- (dash0L1)
        (dash0L2)  -- (end01)
        (start001) -- (dash00L1)
        (dash00L2) -- (end001)
        (start0001) -- (dash000L1)
        (dash000L2) -- (end0001)
        (start00001) -- (dash0000L1)
        (dash0000L2) -- (end00001)
        (start11) -- (dash1L1)
        (dash1L2) -- (end11)
        (start101) -- (dash10L1)
        (dash10L2) -- (end101)
        (start1001) -- (dash100L1)
        (dash100L2) -- (end1001)
        (start10001) -- (dash1000L1)
        (dash1000L2) -- (end10001)
        ;
        \draw[thick,dotted]
        (startM1) -- (endM1)
        (startML1) -- (endML1)
        (dash0L1) -- (dash0L2)
        (dash00L1) -- (dash00L2)
        (dash000L1) -- (dash000L2)
        (dash0000L1) -- (dash0000L2)
        (dash1L1) -- (dash1L2)
        (dash10L1) -- (dash10L2)
        (dash100L1) -- (dash100L2)
        (dash1000L1) -- (dash1000L2)
        ;
        \draw[thick,shorten >=-4pt,shorten <=-4pt](X01)--(X001);
        \draw[thick,shorten >=-4pt,shorten <=-4pt](X0001)--(X00001);
        \draw[thick,shorten >=-4pt,shorten <=-4pt](X002)--(X00002);
        \draw[thick,shorten >=-4pt,shorten <=-4pt](X11)--(X101);
        \draw[thick,shorten >=-4pt,shorten <=-4pt](X1001)--(X10001);
        \draw[thick,shorten >=-4pt,shorten <=-4pt](X102)--(X10002);
        \draw[thick,shorten >=-4pt,shorten <=-4pt](X00003)+(0em,0.1em)--(midLeft)+(0em,0.1em);
        \draw[thick,shorten >=-4pt,shorten <=-4pt](X10003)+(0em,0.1em)--(midLeftL)+(0em,0.1em);
        \draw[thick,shorten >=-4pt,shorten <=-4pt](X10004)+(0em,0.1em)--(midRightL)+(0em,0.1em);
      \end{pgfonlayer}
    \end{tikzpicture}
    \\
  \end{tabular}
  \end{center}
  \caption{
    \label{fig:binary_tree_circuit_type_1b}
    Qutrit decomposition of the multi-controlled Toffoli gate, using a
    binary-tree topology and circuit of Type-B.
  }
\end{figure}

When using Type-B circuits, an implementation of the binary tree with qutrits is straightforward, and is shown in Figure~\ref{fig:binary_tree_circuit_type_1b}. A binary tree differs from the linear chain in that independent sub-trees do not influence one another at lower layers. We might therefore assume that the error associated with higher-index states must have the linear dependence seen for the Type-B linear chain, since active gates are not tethered to any particular point. This is true, though as we shall see the precise scaling makes larger modified clusters less likely than in the linear case.

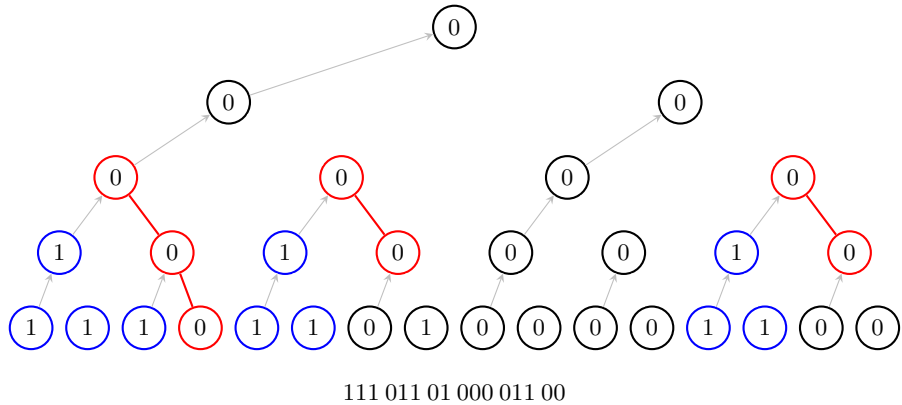
\begin{figure}[ht!]
  \begin{center}
    \begin{tikzpicture}[font=\small,thick,level distance=1cm]
    \node[black,draw=black,circle,fill=white,minimum width=0.5cm]
    (00) {$0$} [sibling distance = 6.0cm]
      child {node[black,draw=black,circle,fill=white,minimum width=0.5cm]
        (10) {$0$} [sibling distance = 3.0cm] edge from parent [white]
        child {node[black,draw=red,circle,fill=white,minimum width=0.5cm]
          (20) {$0$} [sibling distance = 1.5cm] edge from parent [white]
          child {node[black,draw=blue,circle,fill=white,minimum width=0.5cm]
            (30) {$1$} [sibling distance = 0.75cm] edge from parent [white]
            child {node[black,draw=blue,circle,fill=white,minimum width=0.5cm]
              (40) {$1$} edge from parent [white]
            }
            child {node[black,draw=blue,circle,fill=white,minimum width=0.5cm]
              (41) {$1$} edge from parent [white]
            }
          }
          child {node[black,draw=red,circle,fill=white,minimum width=0.5cm]
            (31) {$0$} [sibling distance = 0.75cm] edge from parent [white]
            child {node[black,draw=blue,circle,fill=white,minimum width=0.5cm]
              (42) {$1$} edge from parent [white]
            }
            child {node[black,draw=red,circle,fill=white,minimum width=0.5cm]
              (43) {$0$} edge from parent [white]
            }
          }
        }
        child {node[black,draw=red,circle,fill=white,minimum width=0.5cm]
          (21) {$0$} [sibling distance = 1.5cm] edge from parent [white]
          child {node[black,draw=blue,circle,fill=white,minimum width=0.5cm]
            (32) {$1$} [sibling distance = 0.75cm] edge from parent [white]
            child {node[black,draw=blue,circle,fill=white,minimum width=0.5cm]
              (44) {$1$} edge from parent [white]
            }
            child {node[black,draw=blue,circle,fill=white,minimum width=0.5cm]
              (45) {$1$} edge from parent [white]
            }
          }
          child {node[black,draw=red,circle,fill=white,minimum width=0.5cm]
            (33) {$0$} [sibling distance = 0.75cm] edge from parent [white]
            child {node[black,draw=black,circle,fill=white,minimum width=0.5cm]
              (46) {$0$} edge from parent [white]
            }
            child {node[black,draw=black,circle,fill=white,minimum width=0.5cm]
              (47) {$1$} edge from parent [white]
            }
          }
        }
      }
      child {node[black,draw=black,circle,fill=white,minimum width=0.5cm]
        (11) {$0$} [sibling distance = 3.0cm] edge from parent [white]
        child {node[black,draw=black,circle,fill=white,minimum width=0.5cm]
          (22) {$0$} [sibling distance = 1.5cm] edge from parent [white]
          child {node[black,draw=black,circle,fill=white,minimum width=0.5cm]
            (34) {$0$} [sibling distance = 0.75cm] edge from parent [white]
            child {node[black,draw=black,circle,fill=white,minimum width=0.5cm]
              (48) {$0$} edge from parent [white]
            }
            child {node[black,draw=black,circle,fill=white,minimum width=0.5cm]
              (49) {$0$} edge from parent [white]
            }
          }
          child {node[black,draw=black,circle,fill=white,minimum width=0.5cm]
            (35) {$0$} [sibling distance = 0.75cm] edge from parent [white]
            child {node[black,draw=black,circle,fill=white,minimum width=0.5cm]
              (410) {$0$} edge from parent [white]
            }
            child {node[black,draw=black,circle,fill=white,minimum width=0.5cm]
              (411) {$0$} edge from parent [white]
            }
          }
        }
        child {node[black,draw=red,circle,fill=white,minimum width=0.5cm]
          (23) {$0$} [sibling distance = 1.5cm] edge from parent [white]
          child {node[black,draw=blue,circle,fill=white,minimum width=0.5cm]
            (36) {$1$} [sibling distance = 0.75cm] edge from parent [white]
            child {node[black,draw=blue,circle,fill=white,minimum width=0.5cm]
              (412) {$1$} edge from parent [white]
            }
            child {node[black,draw=blue,circle,fill=white,minimum width=0.5cm]
              (413) {$1$} edge from parent [white]
            }
          }
          child {node[black,draw=red,circle,fill=white,minimum width=0.5cm]
            (37) {$0$} [sibling distance = 0.75cm] edge from parent [white]
            child {node[black,draw=black,circle,fill=white,minimum width=0.5cm]
              (414) {$0$} edge from parent [white]
            }
            child {node[black,draw=black,circle,fill=white,minimum width=0.5cm]
              (415) {$0$} edge from parent [white]
            }
          }
        }
      }
    ;
    \draw[thin,lightgray,stealth-] (00) -- (10);
    \draw[thin,lightgray,stealth-] (10) -- (20);
    \draw[thin,lightgray,stealth-] (20) -- (30);
    \draw[thin,lightgray,stealth-] (30) -- (40);
    \draw[thin,lightgray,stealth-] (21) -- (32);
    \draw[thin,lightgray,stealth-] (32) -- (44);
    \draw[thin,lightgray,stealth-] (31) -- (42);
    \draw[thin,lightgray,stealth-] (33) -- (46);
    \draw[thin,lightgray,stealth-] (11) -- (22);
    \draw[thin,lightgray,stealth-] (22) -- (34);
    \draw[thin,lightgray,stealth-] (23) -- (36);
    \draw[thin,lightgray,stealth-] (34) -- (48);
    \draw[thin,lightgray,stealth-] (35) -- (410);
    \draw[thin,lightgray,stealth-] (36) -- (412);
    \draw[thin,lightgray,stealth-] (37) -- (414);
    \draw[red,thick]
    (20) -- (31)
    (31) -- (43)
    (21) -- (33)
    (23) -- (37)
    ;
    \end{tikzpicture}
    \begin{align*}
      111\: 011\: 01\: 000\: 011\: 00
    \end{align*}
    \caption{\label{fig:binary_tree_entanglement_propagation_circuit_type_1b}
      Propagation of active gates and modified bits in a Type-B circuit with a binary
      tree topology, for the specific example of the bit-string
      $111\: 011\: 01\: 000\: 011\: 00$, where the left-hand bit in each pair moves
      to the next layer.
      Highlighted in red are the modified bits and their connections. Highlighted in
      blue are the $1$-cluster sub-trees necessary for active gate propagation.
      Since nodes at higher layers are generated from those at lower layers, it is
      clear that these sub-trees correspond to $1$-runs in the original bit-string,
      at specific positions partitioned by the tree structure.
    }
  \end{center}
\end{figure}

At the lowest layer, for $n$ total qutrits, two-qutrit gates will be active in any of the independent $n/2$ pairs where the first qutrit in the pair is in the \ket{0} state and the second is in the \ket{1} state. This configuration occurs with probability $1/4$, so at the conclusion of the first step, for a uniform initial superposition, we expect an average population of $n/8$ $\vert2\rangle$ states. At the second layer the situation repeats, as shown in
Figure~\ref{fig:binary_tree_entanglement_propagation_circuit_type_1b}.

Working up recursively, we see that:
\begin{enumerate}
  \item
  For a tree of depth two (a $2$-bit string), there is one configuration ($10$) that results in an active gate and two-bit modification.
  \item
  Moving to depth three adds one configuration in which two active gates modify three bits (the configuration on the far left of
  Figure~\ref{fig:binary_tree_entanglement_propagation_circuit_type_1b}), two configurations in which the top node is one of two modified bits (the left sub-tree must consist of all ones, while the right sub-tree can take any two-out-of-four configurations resulting in a zero node without an active gate). In addition we have a single configuration from each sub-tree that modifies two bits without involving the top node.
  \item
  In general, in moving from depth $N$ to depth $N+1$ we observe:
  \begin{enumerate}
    \item
    The numbers from the depth $N$ sub-trees are included twice over, allowing for all additional sub-configurations, to account for those cases in which the top node is not connected to any of its immediate predecessors.
    \item
    There is one case in which the top node is connected to all of its $N$ immediate predecessors.
    \item
    The top node is connected to $k$ of its immediate predecessors ($0 < k < N$) in $2^{2^{N-k-1}}\left(2^{2^{N-k-1}}-1\right)$ configurations, within which additional separate connections may exist within each of two, depth-$(N-k)$ sub-trees.
  \end{enumerate}
  \item
  Let $R^{(N)}_{k}$ represent the number of disjoint paths of exactly length $k>0$ that can be found across all configurations in a tree of depth-$N$. We have:
  \begin{align}
    R^{(N+1)}_{k>N} &= 0
    \nonumber\\
    R^{(N+1)}_{N} &= 1
    \nonumber\\
    R^{(N+1)}_{k<N} &= 2^{2^{N-1}+1} R^{(N)}_{k} + \left(2^{2^{N-k-1}+1}-1\right) R^{(N-k)}_{k} + 2^{2^{N-k-1}}\left(2^{2^{N-k-1}}-1\right)
    ,
    \label{eq:binary_tree_1b_recurrence_relation}
  \end{align}
  where $2^{2^{N-1}+1} R^{(N)}_{k}$ represents $2^{2^{N-1}} R^{(N)}_{k}$ points appearing in a left sub-tree as we cycle through all $2^{2^{N-1}}$ configurations of the right sub-tree (of depth-$N$), and this is doubled to incorporate those in the right sub-tree via symmetry (there is no double-counting here because we only count points from one sub-tree in each component).
\end{enumerate}

Since
\begin{align}
  2^{2^{N-j}}
  &=
  \left(2^{2^{N}}\right)^{\frac{1}{2^{j}}}
  ,
\end{align}
we can re-express the recursion relation as
\begin{align}
  R^{(N+1)}_{k<N}
  &=
  2 \left(2^{2^{N}}\right)^{1/2} R^{(N)}_{k} + \left(2 \left(2^{2^{N}}\right)^{1/2^{k+1}}-1\right) R^{(N-k)}_{k} + \left(2^{2^{N}}\right)^{1/2^{k+1}}\left(\left(2^{2^{N}}\right)^{1/2^{k+1}}-1\right)
  ,
  \label{eq:binary_tree_1b_recurrence_relation_sqrts}
\end{align}
making it clear that the dominant term as $N$ becomes large is $2 \left(2^{2^{N}}\right)^{1/2} R^{(N)}_{k}$. Now, the total number of possible configurations at depth $N+1$ is $2^{2^{N}}$, so that the density scales as $2 \left(2^{2^{N}}\right)^{-1/2} R^{(N)}_{k}$. The dominant term in $R^{(N)}_{k}$ is $2 \left(2^{2^{N}}\right)^{1/4} R^{(N-1)}_{k}$, and continuing down the dominant track recursively we have
\begin{align}
  \frac{R^{(N)}_{k}}{2^{2^{N}}}
  &\sim 2 \left(2^{2^{N}}\right)^{-1/2} R^{(N)}_{k}
  \nonumber\\
  &\sim 4 \left(2^{2^{N}}\right)^{-1/4} R^{(N-1)}_{k}
  \nonumber\\
  &\sim 8 \left(2^{2^{N}}\right)^{-1/8} R^{(N-2)}_{k}
  \nonumber\\
  &\rightarrow
  2^{N-k} \left(2^{2^{N}}\right)^{-1/2^{N-k}}
  \nonumber\\
  &= 2^{N} 2^{-2^{k}-k}
  \nonumber\\
  &= n 2^{-2^{k}-k}
  ,
\end{align}
where $n$ is the total number of bits.

We make two observations:
\begin{enumerate}
  \item
  The number of modified clusters of a given size, per configuration, increases linearly with the number of qudits, just as for the Type-B linear circuit.
  \item
  Larger modified clusters are suppressed according to a double-exponential, rather than a simple exponential as in the linear circuit; log-depth chains require linear-length $1$-cluster sequences.
\end{enumerate}

\subsection{Type-A Circuits}
\label{sub:type_1a_circuits}

In the linear version of the Type-A circuit, some information about active gates was stored in the third state of each successive target. Similarly, for all Type-B circuits, this information is stored in the control-qudits, leaving space in the targets free for subsequent steps. However, when we attempt to implement a Type-A binary tree structure we find that it is not possible to assign a qutrit as a target in multiple layers without increasing its dimension.

There therefore appear to be two methods for implementing Type-A binary trees:
\begin{enumerate}
  \item
  Increase the dimension of each qudit according to the number of times it is assigned as a target, along the lines of the star-like network topologies
  \cite{ralph_efficient_2007,lanyon_simplifying_2009}
  discussed
  in Section~III~D of the main text.
  \item
  Use two-controlled Toffoli gates \cite{gokhale_asymptotic_2019}, so that outcomes from multiple branches can be combined in a fresh target.
\end{enumerate}
Figure~$4$ of the main text shows a representative circuit implementation for each method, which we'll refer to as the `re-use' and `two-control' methods respectively.

\subsubsection{Type-A `re-use' method}

For the Type-B network of
Figure~\ref{fig:binary_tree_entanglement_propagation_circuit_type_1b},
activation at each layer was dependent on a 1-cluster for the left sub-tree in the layer below, but allowed any other (non-$1$-cluster) configuration for the right sub-tree.
For the Type-A `re-use' network of main text
Figure~$4$,
by contrast, activation at each layer requires prior activation at the top node of each sub-tree in the layer below. This occurs for only one configuration; activation at higher layers
is exponentially less likely than for the Type-B network.
However, for the Type-A `re-use' network, all qubits under an active node will themselves have been involved in an active gate, so that the size of a modified cluster grows exponentially in the index of its topmost layer. In the Type-B network the layer index bounded the size of the modified cluster logarithmically.

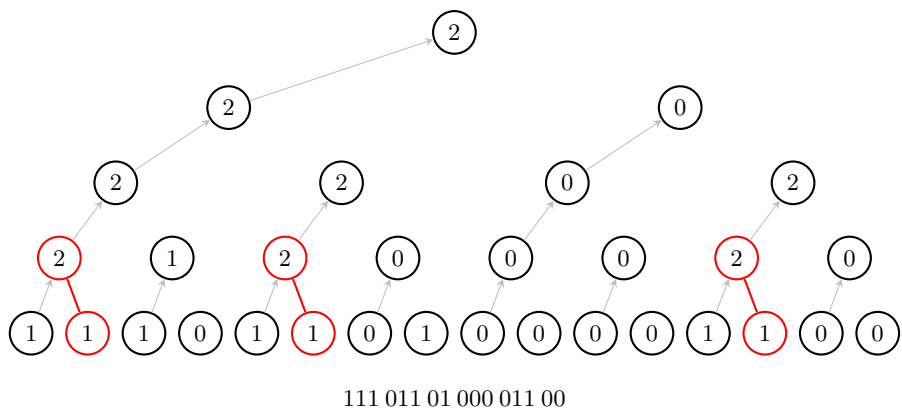
\begin{figure}[ht!]
  \begin{center}
    \begin{tikzpicture}[font=\small,thick,level distance=1cm]
    \node[black,draw=black,circle,fill=white,minimum width=0.5cm]
    (00) {$2$} [sibling distance = 6.0cm]
      child {node[black,draw=black,circle,fill=white,minimum width=0.5cm]
        (10) {$2$} [sibling distance = 3.0cm] edge from parent [white]
        child {node[black,draw=black,circle,fill=white,minimum width=0.5cm]
          (20) {$2$} [sibling distance = 1.5cm] edge from parent [white]
          child {node[black,draw=red,circle,fill=white,minimum width=0.5cm]
            (30) {$2$} [sibling distance = 0.75cm] edge from parent [white]
            child {node[black,draw=black,circle,fill=white,minimum width=0.5cm]
              (40) {$1$} edge from parent [white]
            }
            child {node[black,draw=red,circle,fill=white,minimum width=0.5cm]
              (41) {$1$} edge from parent [white]
            }
          }
          child {node[black,draw=black,circle,fill=white,minimum width=0.5cm]
            (31) {$1$} [sibling distance = 0.75cm] edge from parent [white]
            child {node[black,draw=black,circle,fill=white,minimum width=0.5cm]
              (42) {$1$} edge from parent [white]
            }
            child {node[black,draw=black,circle,fill=white,minimum width=0.5cm]
              (43) {$0$} edge from parent [white]
            }
          }
        }
        child {node[black,draw=black,circle,fill=white,minimum width=0.5cm]
          (21) {$2$} [sibling distance = 1.5cm] edge from parent [white]
          child {node[black,draw=red,circle,fill=white,minimum width=0.5cm]
            (32) {$2$} [sibling distance = 0.75cm] edge from parent [white]
            child {node[black,draw=black,circle,fill=white,minimum width=0.5cm]
              (44) {$1$} edge from parent [white]
            }
            child {node[black,draw=red,circle,fill=white,minimum width=0.5cm]
              (45) {$1$} edge from parent [white]
            }
          }
          child {node[black,draw=black,circle,fill=white,minimum width=0.5cm]
            (33) {$0$} [sibling distance = 0.75cm] edge from parent [white]
            child {node[black,draw=black,circle,fill=white,minimum width=0.5cm]
              (46) {$0$} edge from parent [white]
            }
            child {node[black,draw=black,circle,fill=white,minimum width=0.5cm]
              (47) {$1$} edge from parent [white]
            }
          }
        }
      }
      child {node[black,draw=black,circle,fill=white,minimum width=0.5cm]
        (11) {$0$} [sibling distance = 3.0cm] edge from parent [white]
        child {node[black,draw=black,circle,fill=white,minimum width=0.5cm]
          (22) {$0$} [sibling distance = 1.5cm] edge from parent [white]
          child {node[black,draw=black,circle,fill=white,minimum width=0.5cm]
            (34) {$0$} [sibling distance = 0.75cm] edge from parent [white]
            child {node[black,draw=black,circle,fill=white,minimum width=0.5cm]
              (48) {$0$} edge from parent [white]
            }
            child {node[black,draw=black,circle,fill=white,minimum width=0.5cm]
              (49) {$0$} edge from parent [white]
            }
          }
          child {node[black,draw=black,circle,fill=white,minimum width=0.5cm]
            (35) {$0$} [sibling distance = 0.75cm] edge from parent [white]
            child {node[black,draw=black,circle,fill=white,minimum width=0.5cm]
              (410) {$0$} edge from parent [white]
            }
            child {node[black,draw=black,circle,fill=white,minimum width=0.5cm]
              (411) {$0$} edge from parent [white]
            }
          }
        }
        child {node[black,draw=black,circle,fill=white,minimum width=0.5cm]
          (23) {$2$} [sibling distance = 1.5cm] edge from parent [white]
          child {node[black,draw=red,circle,fill=white,minimum width=0.5cm]
            (36) {$2$} [sibling distance = 0.75cm] edge from parent [white]
            child {node[black,draw=black,circle,fill=white,minimum width=0.5cm]
              (412) {$1$} edge from parent [white]
            }
            child {node[black,draw=red,circle,fill=white,minimum width=0.5cm]
              (413) {$1$} edge from parent [white]
            }
          }
          child {node[black,draw=black,circle,fill=white,minimum width=0.5cm]
            (37) {$0$} [sibling distance = 0.75cm] edge from parent [white]
            child {node[black,draw=black,circle,fill=white,minimum width=0.5cm]
              (414) {$0$} edge from parent [white]
            }
            child {node[black,draw=black,circle,fill=white,minimum width=0.5cm]
              (415) {$0$} edge from parent [white]
            }
          }
        }
      }
    ;
    \draw[thin,lightgray,stealth-] (00) -- (10);
    \draw[thin,lightgray,stealth-] (10) -- (20);
    \draw[thin,lightgray,stealth-] (20) -- (30);
    \draw[thin,lightgray,stealth-] (30) -- (40);
    \draw[thin,lightgray,stealth-] (21) -- (32);
    \draw[thin,lightgray,stealth-] (32) -- (44);
    \draw[thin,lightgray,stealth-] (31) -- (42);
    \draw[thin,lightgray,stealth-] (33) -- (46);
    \draw[thin,lightgray,stealth-] (11) -- (22);
    \draw[thin,lightgray,stealth-] (22) -- (34);
    \draw[thin,lightgray,stealth-] (23) -- (36);
    \draw[thin,lightgray,stealth-] (34) -- (48);
    \draw[thin,lightgray,stealth-] (35) -- (410);
    \draw[thin,lightgray,stealth-] (36) -- (412);
    \draw[thin,lightgray,stealth-] (37) -- (414);
    \draw[red,thick]
    (41) -- (30)
    (45) -- (32)
    (413) -- (36)
    ;
    \end{tikzpicture}
    \begin{align*}
      111\: 011\: 01\: 000\: 011\: 00
    \end{align*}
    \caption{\label{fig:binary_tree_entanglement_propagation_circuit_type_1a_reuse}
      Propagation of active gates and modified bits in a Type-A circuit with a binary
      tree topology, applying the `re-use' method (see main text
      Figure~$4$),
      for the specific example of the bit-string
      $111\: 011\: 01\: 000\: 011\: 00$,
      where the left-hand bit in each pair moves to the next layer.
      Highlighted in red are the modified bits and their connections.
    }
  \end{center}
\end{figure}

For the Type-A `re-use' circuit in a tree of depth-$N$,
at the $k$th layer there are $2^{k-1}$ sub-trees
(the root node is at layer $1$).
Each of these sub-trees can independently support a single modified cluster
involving its top node, in one of its $2^{2^{N-k}}$ configurations.
This cluster incorporates all $2^{N-k}$ qudits in the sub-tree.
Therefore, in all configurations of the network there are
\begin{align}
  \underbrace{2^{k-1}}_{\text{sub-trees}}
  \cdot
  \left.
  \underbrace{\left[
    2^{2^{N-k}}
  \right]}
  _{\text{Configs.}}
  \right.^{\underbrace{\left(2^{k-1}-1\right)}_{\text{sub-trees }- 1}}
\end{align}
instances of a modified cluster of size $2^{N-k}$ (involving this many qudits).
Expressing the number of qudits involved in the modified cluster as $r$ and the total number of qudits as $n=2^{N-1}$, this is
\begin{align}
  \frac{n}{r} 2^{n} 2^{-r}
  .
\end{align}
In the total network there are $2^{n}$ configurations, so that the number of instances per configuration is
\begin{align}
  \frac{n}{r} 2^{-r}
  .
\end{align}
Once again, we have an exponential decay in $r$ (not a double-exponential this time), and a linear increase with $n$.

\subsubsection{Type-A `two-control' method}

The `two-control' method has counting statistics very similar to the `re-use' method, except that the branching ratio of the circuit is $3$, and the central branch at each point consists of only a single node drawn directly from the input.
Here a depth-$N$ tree contains $2^{N}-1$ qudits, or $2^{2^{N}-1}$ configurations.

At layer $k<N$ (the root node is at layer $1$), there are $2^{k-1}$ non-trivial sub-trees (excluding single-node branches). For each non-trivial sub-tree there are $2^{2^{N-k+1}-1}$ configurations. For only one configuration, involving all $2^{N-k+1}-1$ qudits in the sub-tree, is the top node of the sub-tree active.

Therefore, in all configurations of the network there are
\begin{align}
  \underbrace{2^{k-1}}_{\text{sub-trees}}
  \cdot
  \left.
  \underbrace{\left[
    2^{2^{N-k+1}-1}
  \right]}
  _{\text{Configs.}}
  \right.^{\underbrace{\left(2^{k-1}-1\right)}_{\text{sub-trees }- 1}}
\end{align}
instances of a modified cluster of size $2^{N-k+1}-1$ (involving this many qudits).

Expressing the number of qudits involved in the modified cluster as $r$ and the total number of qudits as $n=2^{N}-1$, this is
\begin{align}
  \frac{n+1}{r+1}
  2^{
    n - r
    - \frac{n+1}{r+1}
  }
  ,
\end{align}
giving the number of instances per configuration as
\begin{align}
  \frac{n+1}{r+1}
  2^{
    - r
    - \frac{n+1}{r+1}
  }
  .
\end{align}
Once again we have a linear dependence on $n$ and an exponential decay with $r$, but in this case there is also an exponential decay with $\frac{n}{r+1}$:
There are single-node branches introduced with each higher layer that cannot contribute to size-$r$ modified clusters at lower layers, but which still enlarge the number of configurations exponentially.
The network structure in this case is a kind of hybrid between a pure binary tree with independent leaves and the Type-A linear chain with all clusters tethered to qudits at the one boundary of the circuit, those involved in the initial gates.
The additional exponential decay bounds the size of the error per configuration as the circuit size increases, so that once again a gate-level error model is appropriate.

We could consider the temporary entanglement generated inside each
three-qudit gate, but since these are of constant size $3$ they simply correspond to
the three circuits shown in Figure~$3$ of the main text,
and so fall under the prior discussion on linear and star-like networks.

\begin{figure}[ht!]
  \begin{center}
    \begin{tikzpicture}[font=\small,thick,level distance=1cm]
    \node[black,draw=black,circle,fill=white,minimum width=0.5cm]
      (10) {$1$} [sibling distance = 3.0cm]
      child {node[black,draw=black,circle,fill=white,minimum width=0.5cm]
        (20) {$0$} [sibling distance = 1.5cm] edge from parent [white]
        child {node[black,draw=red,circle,fill=white,minimum width=0.5cm]
          (30) {$2$} [sibling distance = 0.75cm] edge from parent [white]
          child {node[black,draw=red,circle,fill=white,minimum width=0.5cm]
            (40) {$1$} [sibling distance = 0.55cm] edge from parent [white]
          }
          child {node[black,draw=black,circle,fill=white,minimum width=0.5cm]
            (41) {$1$} [sibling distance = 0.55cm] edge from parent [white]
          }
          child {node[black,draw=red,circle,fill=white,minimum width=0.5cm]
            (42) {$1$} [sibling distance = 0.55cm] edge from parent [white]
          }
        }
        child {node[black,draw=black,circle,fill=white,minimum width=0.5cm]
          (31) {$0$} [sibling distance = 0.75cm] edge from parent [white]
          child {node[black,draw=black,circle,fill=white,minimum width=0.5cm]
            (43) {$0$} [sibling distance = 0.75cm] edge from parent [white]
          }
        }
        child {node[black,draw=black,circle,fill=white,minimum width=0.5cm]
          (32) {$1$} [sibling distance = 0.75cm] edge from parent [white]
          child {node[black,draw=black,circle,fill=white,minimum width=0.5cm]
            (44) {$1$} [sibling distance = 0.75cm] edge from parent [white]
          }
          child {node[black,draw=black,circle,fill=white,minimum width=0.5cm]
            (45) {$1$} [sibling distance = 0.75cm] edge from parent [white]
          }
          child {node[black,draw=black,circle,fill=white,minimum width=0.5cm]
            (46) {$0$} [sibling distance = 0.75cm] edge from parent [white]
          }
        }
      }
      child {node[black,draw=black,circle,fill=white,minimum width=0.5cm]
        (21) {$1$} [sibling distance = 1.25cm] edge from parent [white]
        child {node[black,draw=black,circle,fill=white,minimum width=0.5cm]
          (33) {$1$} [sibling distance = 0.75cm] edge from parent [white]
          child {node[black,draw=black,circle,fill=white,minimum width=0.5cm]
            (47) {$1$} [sibling distance = 0.75cm] edge from parent [white]
          }
        }
      }
      child {node[black,draw=black,circle,fill=white,minimum width=0.5cm]
        (22) {$0$} [sibling distance = 1.5cm] edge from parent [white]
        child {node[black,draw=black,circle,fill=white,minimum width=0.5cm]
          (34) {$0$} [sibling distance = 0.75cm] edge from parent [white]
          child {node[black,draw=black,circle,fill=white,minimum width=0.5cm]
            (48) {$0$} [sibling distance = 0.75cm] edge from parent [white]
          }
          child {node[black,draw=black,circle,fill=white,minimum width=0.5cm]
            (49) {$0$} [sibling distance = 0.75cm] edge from parent [white]
          }
          child {node[black,draw=black,circle,fill=white,minimum width=0.5cm]
            (410) {$0$} [sibling distance = 0.75cm] edge from parent [white]
          }
        }
        child {node[black,draw=black,circle,fill=white,minimum width=0.5cm]
          (35) {$0$} [sibling distance = 0.75cm] edge from parent [white]
          child {node[black,draw=black,circle,fill=white,minimum width=0.5cm]
            (411) {$0$} [sibling distance = 0.75cm] edge from parent [white]
          }
        }
        child {node[black,draw=black,circle,fill=white,minimum width=0.5cm]
          (36) {$1$} [sibling distance = 0.75cm] edge from parent [white]
          child {node[black,draw=black,circle,fill=white,minimum width=0.5cm]
            (412) {$1$} [sibling distance = 0.75cm] edge from parent [white]
          }
          child {node[black,draw=black,circle,fill=white,minimum width=0.5cm]
            (413) {$1$} [sibling distance = 0.75cm] edge from parent [white]
          }
          child {node[black,draw=black,circle,fill=white,minimum width=0.5cm]
            (414) {$0$} [sibling distance = 0.75cm] edge from parent [white]
          }
        }
      }
    ;
    \draw[thin,lightgray,stealth-] (30) -- (41);
    \draw[thin,lightgray,stealth-] (31) -- (43);
    \draw[thin,lightgray,stealth-] (20) -- (31);
    \draw[thin,lightgray,stealth-] (32) -- (45);
    \draw[thin,lightgray,stealth-] (10) -- (21);
    \draw[thin,lightgray,stealth-] (21) -- (33);
    \draw[thin,lightgray,stealth-] (33) -- (47);
    \draw[thin,lightgray,stealth-] (34) -- (49);
    \draw[thin,lightgray,stealth-] (22) -- (35);
    \draw[thin,lightgray,stealth-] (35) -- (411);
    \draw[thin,lightgray,stealth-] (36) -- (413);
    \draw[red,thick]
    (40)--(30)
    (42)--(30)
    ;
    \end{tikzpicture}
    \begin{align*}
      111\: 011\: 01\: 000\: 011\: 0
    \end{align*}
    \caption{\label{fig:binary_tree_entanglement_propagation_circuit_type_1a_twocontrol}
      Propagation of active gates and modified bits in a Type-A circuit, applying
      the `two-control' method (see main text
      Figure~$4$),
      for the specific example of the bit-string
      $111\: 011\: 01\: 000\: 011\: 0$,
      where the central bit in each triplet moves to the next layer.
      Highlighted in red are the modified bits and their connections.
      Here the bit-string is one element shorter than at other
      places in this document, so that the number of bits divides evenly
      into the tree structure.
    }
  \end{center}
\end{figure}